\documentstyle[prd,aps]{revtex}

\input epsf
\def\gtrsim{\mathrel{\hbox{\rlap{\hbox{\lower4pt\hbox{$\sim$}}}\hbox{$>$}}}}

\begin{document}

\draft
\title{Post-Newtonian SPH calculations of binary neutron star 
coalescence. III. Irrotational systems and gravitational wave spectra}
\author{Joshua A.\ Faber, Frederic A.\ Rasio}
\address{Department of Physics and Astronomy, 
 Northwestern University, Evanston, IL 60208}

\date{\today}

\maketitle

\begin{abstract}
Gravitational wave (GW) signals from coalescing binary neutron stars
may soon become detectable by laser-interferometer detectors.
Using our new post-Newtonian (PN) smoothed particle hydrodynamics
(SPH) code, we have studied numerically the mergers of neutron star
binaries with irrotational initial configurations. 
These are the most physically realistic initial conditions
just prior to merger, 
since the neutron stars in these systems are expected to
be spinning slowly at large separation, and the
viscosity of neutron star matter is too small for
tidal synchronization to be effective at small separation.
However, the large shear that develops during the merger makes
irrotational systems particularly difficult to study numerically in 3D.
In addition, in PN gravity, accurate irrotational initial 
conditions are much more difficult to construct numerically than 
corotating initial conditions.
Here we describe a new method for constructing numerically accurate 
initial conditions for irrotational binary systems with 
circular orbits in PN gravity. We then compute
the 3D hydrodynamic evolution of these systems until the
two stars have completely merged, and we determine the corresponding
GW signals. We present results for 
systems with different binary mass ratios, and for neutron 
stars represented by polytropes with $\Gamma=2$ or $\Gamma=3$.
Compared to mergers of corotating binaries, we find that
irrotational binary mergers produce similar peak GW
luminosities, but they shed almost no
mass at all to large distances. The dependence of the GW signal on
numerical resolution for calculations
performed with $N\gtrsim 10^5$ SPH particles is extremely weak, and we find
excellent agreement between runs utilizing $N=10^5$ and $N=10^6$ SPH
particles (the largest SPH calculation ever performed
to study such irrotational binary mergers).  
We also compute GW energy spectra based on all calculations
reported here and in our previous works.  We find that PN
effects lead to clearly identifiable features in the GW energy
spectrum of binary neutron star mergers, which may yield 
important information about the nuclear 
equation of state at extreme densities.
\end{abstract}

\section{Introduction and Motivation}\label{sec:intro}

Coalescing neutron star (NS) binaries are likely to be one of the most
important sources of gravitational radiation for the ground-based
laser-interferometer detectors in LIGO
\cite{1}, VIRGO \cite{2}, GEO600 \cite{3}, and TAMA \cite{4}.
These interferometers are most sensitive to GW signals in the
frequency range from about $10\,{\rm Hz}$ to $300\,{\rm Hz}$, 
which corresponds to the last several
thousand orbits of the inspiral.  During this period, the binary
orbit is decaying very slowly, with the separation 
$r(t)$ and phase $\phi(t)$ following the standard
theoretical treatment for the inspiral of two point masses 
(see, e.g., \cite{Insp}). 
Theoretical templates for the corresponding quasi-periodic GW signals
covering an appropriate range of values for the
NS masses, as well as orbital phases and inclination angles, can be
calculated to great precision (see, e.g., \cite{Tem} and references
therein), and template matching techniques can therefore be used to
extract signals from noisy interferometer data.
When the binary separation $r(t)$ has decreased all the way down to a
few NS radii, the system becomes dynamically unstable \cite{LRS} and the two
stars merge hydrodynamically in $\sim1\,$ms.
The characteristic GW frequency of the final burst-like signal
is $\gtrsim 1\,{\rm kHz}$, outside the range accessible by current broadband 
detectors.
These GW signals may become detectable, however, by the use of signal 
recycling techniques,
which provide increased sensitivity in a narrow frequency band \cite{SR}.
These techniques are now being tested at GEO600, and will be used by the
next generation of ground-based interferometers.  
Point-mass inspiral templates break down during the final few orbits.
Instead, 3D numerical hydrodynamic calculations are required
to describe the binary merger phase and predict theoretically
the GW signals that
will carry information about the NS equation of state (EOS).

The first hydrodynamic calculations of binary NS mergers in Newtonian gravity 
were performed by Nakamura, Oohara and collaborators using a grid-based, 
Eulerian finite-difference code \cite{ON}.  Rasio and Shapiro (\cite{RS},
hereafter RS) later used Lagrangian SPH calculations to study both the 
stability properties of close NS
binaries and the evolution of dynamically unstable systems to
complete coalescence.  Since then, several groups have performed 
increasingly sophisticated calculations
in Newtonian gravity, exploring the full parameter space of the problem with
either SPH \cite{Zhu,Dav,Ros} or the Eulerian, grid-based piecewise parabolic
method (PPM) \cite{New,Swe,Ruf}, and 
focusing on topics as diverse as the GW energy spectrum
\cite{Zhu}, the production of r-process elements \cite{Ros}, and
the neutrino emission as a possible trigger for gamma-ray bursts
\cite{Ruf}.  Some of these Newtonian calculations have included terms
to model approximately the effects of the gravitational radiation reaction
\cite{Ruf}.

The first calculations to include the lowest-order (1PN)
corrections to Newtonian gravity, as well as the lowest-order dissipative
effects of the gravitational radiation 
reaction (2.5PN) were performed by Shibata, Oohara, and  
Nakamura, using an Eulerian grid-based method \cite{Nak3}.
More recently, the authors (\cite{FR1,FR2}, hereafter Paper~1 and Paper~2,
respectively), as well as Ayal et al.\cite{Ayal},
have performed PN SPH calculations,
using the PN hydrodynamics formalism developed by Blanchet, Damour, and Sch\"afer
(\cite{BDS}, hereafter BDS).  These calculations have revealed that
the addition of 1PN terms can have a significant effect on the 
results of hydrodynamic merger calculations, and on the theoretical
predictions for GW signals.
For example, Paper~1 showed a comparison between two calculations for initially
corotating, equal-mass binary NS systems. In the first
calculation, radiation reaction effects were included, but no 1PN
terms were used, whereas a complete set of both 1PN and 2.5PN terms
were included in the second calculation.  It was found that the inclusion of
1PN terms affected the evolution of the system both prior to merger
and during the merger itself.  The final inspiral rate of the PN binary
just prior to merger was
much more rapid, indicating that the orbit became dynamically unstable
at a greater separation.  Additionally, the GW luminosity
produced by the PN system showed a series of
several peaks, absent from the Newtonian calculation. 

Most previous hydrodynamic calculations of binary NS mergers have
assumed corotating initial conditions, and many modeled the stars
as initially spherical.
However, real binary NS are unlikely to be described well by
such initial conditions. Just prior to contact,
tidal deformations can be quite large and the stars can have
very nonspherical shapes \cite{LRS}. Nevertheless,
because of the very low viscosity of the NS fluid, the tidal
synchronization timescale for coalescing NS binaries is expected to
always be longer than the orbital decay timescale \cite{BC}.  
Therefore, a corotating state is unphysical.
In addition, at large separation, the NS in these systems are 
expected to be spinning slowly (see, e.g., \cite{FK}; observed 
spin periods for radio pulsars in double NS systems are
$\gtrsim 50\,$ms, much longer than their final orbital periods). 
Thus, the fluid in close NS binaries should remain nearly
{\it irrotational\/} (in the inertial frame) and the stars in these
systems can be described approximately
by irrotational Riemann ellipsoids \cite{LRS,BC,LomRS,IRRE}. 
In Paper~2, we showed that nonsynchronized initial conditions can
lead to significant differences in the hydrodynamic evolution of coalescing
binaries, especially in the amount of mass ejected
as a result of the rotational instability that develops
during the merger.

There are many difficulties associated with 
numerical calculations of binary mergers with irrotational initial
conditions, especially in PN gravity.  Foremost of these is the difficulty in
preparing the initial, quasi-equilibrium state of the binary system.  
In synchronized
binaries, the two stars are at rest in a reference frame which corotates with
the system, and thus relaxation techniques can be used to construct accurately
the hydrostatic equilibrium initial state in this corotating frame (see \cite{RS}
and Paper~1). For irrotational systems, no such frame exists in which the
entire fluid would appear to be in hydrostatic equilibrium.
Instead, one must determine self-consistently the initial velocity field
of the fluid in the inertial frame.
Otherwise (e.g., when simple spherical models are used), 
spurious oscillations caused by initial deviations from
equilibrium can lead to numerical errors. This is especially of concern in
PN gravity, where the strength of the gravitational force at any
point in the NS contains terms proportional to the gravitational
potential and pressure at 1PN order. 

Another serious problem for hydrodynamic calculations with irrotational 
initial conditions is the issue of spatial resolution and numerical convergence.  
As was first pointed out in RS2, in a frame corotating with
the binary orbital motion, irrotational stars appear
counterspinning so that, when they first make contact during the
coalescence, a vortex sheet is formed along the interface.
This tangential discontinuity is Kelvin-Helmholtz unstable at 
all wavelengths\cite{Dra}. The sheet is expected to break into a turbulent 
boundary layer which propagates into the fluid and generates
vorticity through dissipation on small scales. 
How well this can be handled by 3D numerical calculations 
with limited spatial resolution is unclear.

The outline of our paper is as follows. 
Section~II presents a summary of 
our numerical methods, including a brief description
of our PN SPH code, and an explanation of the method
used to construct irrotational initial conditions in PN gravity.
Additionally, we give the parameters and assumptions
for all new calculations discussed in this paper. 
Section~III presents our numerical results based on SPH calculations
for several representative binary systems, all with irrotational
initial configurations, but varying mass ratios and NS EOS.
To test our numerical methods, we also study the effects of changing 
the initial binary separation and the numerical resolution.
Section~IV presents GW energy spectra calculated from 
the runs in this and previous papers, as well as a discussion of how
the measurement of spectral features could constrain the NS EOS.
A summary of our PN results and possible directions for further research,
including the possibility of fully relativistic SPH calculations,
are presented in Section~V.

\section{Numerical Methods}

\subsection{Conventions and Basic Parameters} \label{sec:formal}

All our calculations were performed using the post-Newtonian (PN)
SPH code described in detail in Papers~1 and~2.  It is a Lagrangian,
particle-based code, with a treatment of relativistic hydrodynamics
and self-gravity adapted from the PN formalism of BDS.  As in our 
previous work, we use a hybrid 1PN/2.5PN
formalism, in which radiation reaction effects are treated at full
strength (i.e., corresponding to realistic NS parameters), 
but 1PN corrections are scaled down by a factor of about 3
to make them numerically tractable (see below).  We did not perform any
new Newtonian calculations for this paper.  All the Poisson-type 
field equations of the BDS
formalism are solved on grids of size $256^3$, including the space 
for zero-padding, which yields the proper boundary conditions.
Shock heating, which is normally treated
via an SPH artificial viscosity, was ignored, since it plays a
negligible role in binary coalescence, especially for fluids with a 
very stiff EOS.
All runs, except those used to study the effects of numerical
resolution on our results, use $5\times 10^4$ SPH particles per NS 
(i.e., the total number of particles
$N=10^5$), independent of the binary mass ratio $q$. 
The number of SPH particle neighbors is set to $N_N=100$ for all runs
with $N=10^5$. 

Unless indicated otherwise, we use
units such that $G=M=R=1$, where $M$ and $R$ are the mass and radius
of one NS.  In calculations for unequal-mass binaries,
we use the mass and radius of the primary. As in our
previous papers, we compute the gravitational radiation reaction
assuming that the speed of light $c_{2.5PN}=2.5$ in our units, 
which corresponds to a neutron star compactness $GM/Rc^2_{2.5PN}=0.16$.
For a standard NS mass of $1.4\,M_\odot$ this also corresponds to a radius
$R=13\,$km. The unit of frequency (used throughout Sec.~IV) is then
\begin{equation} \label{eq:fdyn}
f_{\rm dyn}\equiv \left(\frac{GM}{R^3}\right)^{1/2}
            = 9.2\,{\rm kHz}\,\left(\frac{M}{1.4\,M_\odot}\right)^{1/2}\,
                              \left(\frac{R}{13\,{\rm km}}\right)^{-3/2}
\end{equation}
In order to keep all 1PN terms sufficiently
small with respect to Newtonian quantities, we calculate them
assuming a somewhat larger value of the speed of light, 
$c_{1PN}=4.47$, which would correspond to NS with 
$GM/Rc^2_{1PN}=0.05$.  

As in our previous work, all calculations in this paper use a simple 
polytropic EOS, i.e., the pressure is given in terms
of the mass-energy density by $P=k\rho_*^{\Gamma}$.  We use $\Gamma=3$ to
represent a typical stiff NS EOS, and $\Gamma=2$ to represent a
somewhat softer EOS. The values of the polytropic constant $k$ for all
our NS models (determined by the mass-radius relation) 
are the same as those used in Paper~2 (see Table 1 therein).  
Models for single NS are taken from Papers~1 and~2.  
It should be noted that models of NS with identical masses and
polytropic constants but different
numbers of SPH particles may have slightly different effective
radii, due to numerical resolution effects near the stellar surface.  

It is not easy to define a meaningful and accurate origin of 
time for merger calculations.
The moment of first contact between the two stars is difficult to
determine accurately, since it involves the smoothing lengths of particles
located near the surface of each NS, where the method is least
accurate.  Therefore, following the convention in our earlier papers,
we set the absolute timescale for each run by defining the moment of
peak GW luminosity to be at $t=20$ in our units.  Many runs therefore
start at $t_0<0$, but this is merely a matter of convention.

\subsection{Constructing Irrotational Initial Configurations}
\label{sec:initconf}

To construct irrotational binary configurations in quasi-equilibrium, 
we start from our relaxed, equilibrium models for single NS (Paper~2,
Sec.~IIA).   
These models are transformed linearly into irrotational triaxial ellipsoids, 
with principal axes taken from the PN models of Lombardi, Rasio, and Shapiro 
for equilibrium NS binaries \cite{LomRS}.  For example, for 
an equal-mass system containing two $\Gamma=3$ polytropes with 
an initial separation $r_0=4.0$, we find from their Table~III that the 
principal axes are given by $a_1/R\simeq 1.02$, $a_2/a_1\simeq 0.96$, and
$a_3/a_1\simeq 0.96$, where $a_1$, $a_2$, and $a_3$ are measured
along the binary axis (x-direction), the orbital
motion (y-direction), and the rotation axis (z-direction),
respectively.  For an initial separation $r_0=3.5$,
we find $a_1/R\simeq 1.05$, $a_2/a_1\simeq 0.93$, and $a_3/a_1\simeq 0.93$. 
For the initial velocity field of the fluid we adopt the simple
form assumed for the internal fluid motion
in irrotational Riemann ellipsoids\cite{LRS},
\begin{eqnarray}
v_x=-\Omega y\left( 1-\frac{2a_1^2}{a_1^2+a_2^2}\right) \label{eq:vx}\\
v_y=\Omega x\left( 1-\frac{2a_2^2}{a_1^2+a_2^2}\right), \label{eq:vy}
\end{eqnarray}
where $\Omega$ is the orbital angular velocity.
It is easy to verify that this initial velocity field has zero
vorticity in the inertial frame.

Calculating self-consistently the correct value of $\Omega$ 
corresponding to a quasi-equilibrium circular orbit proves to be much more
difficult in PN gravity than in Newtonian gravity, especially for 
irrotational configurations.  In Newtonian gravity, the gravitational
force between the two stars is independent of the magnitude of
the tidal deformation of
the bodies to lowest order.  Thus, even if the initial configuration
of matter in the binary is slightly out of equilibrium, the orbit calculated
for the two stars will still be almost perfectly circular.  
In PN gravity, the situation is quite different.  From 
Appendix~A of Paper~1, we see that the gravitational acceleration,in our PN
formalism, denoted there as $F_{grav}$, is calculated as 
\begin{equation}
F_{grav}=(1+\frac{1}{c^2}\left[\frac{3\Gamma-2}{\Gamma-1}
\frac{P}{r_*}-U_*+\frac{3w^2}{2}\right])\nabla U_*+\ldots,
\end{equation}
where $w$ is the
the 1PN-adjusted velocity, and $P$, $r_*$,
and $U_*$ are the pressure, 
rest-mass density, and gravitational
potential, respectively, as defined in the BDS
formalism (note that $U_*$ as defined here is a positive quantity).
The quantity in brackets is the 1PN correction to the gravitational
force, and depends explicitly on the initial thermodynamic state 
of the NS.  
Small oscillations of each star about equilibrium result in errors
when calculating the gravitational force felt by each component of the
binary.  In addition, the gravitational force is affected by the
orbital velocity of the NS.  Thus, when we try to calculate the
orbital velocity from the centripetal accelerations of the respective
NS, as in Eq. (1) of Paper~2,
\begin{equation}
\Omega=\sqrt{\frac{-\dot{v}^{(1)}_x+\dot{v}^{(2)}_x}{2r_0}},
\end{equation}
where $v^{(1)}$ and $v^{(2)}$ are the center-of-mass velocities of the
primary (located initially on the positive x-axis) and the secondary
(located on the negative x-axis), respectively,
we face the problem that the RHS of the equation is itself a function
of $\Omega$.

\begin{figure}
\centering \leavevmode \epsfxsize=6in \epsfbox{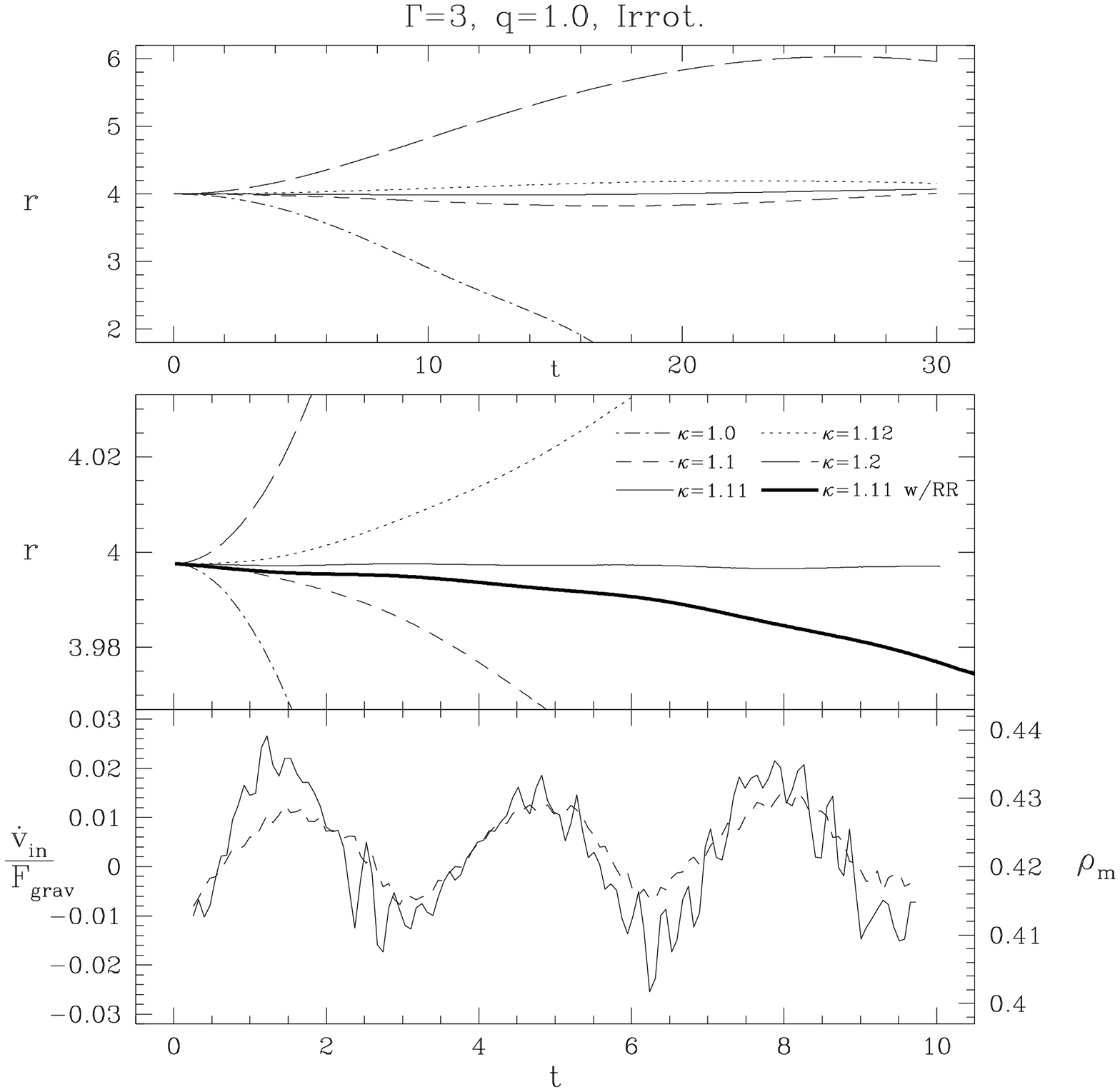}
\caption{The binary separation as a function of time (top panel) 
for purely dynamical runs (without
radiation reaction effects) and $\kappa=1.0$ (dot-dashed line), $kappa=1.1$
(short dashed line), $\kappa=1.11$ (thin solid line), $\kappa=1.12$ (dotted line),
and $\kappa=1.2$ (long dashed line).
The same runs, as well a run with $\kappa=1.11$
including radiation reaction (thick solid line) are shown in the
middle panel, demonstrating the magnitude of the radiation reaction
effects.  We note a very small amplitude radial oscillation in the run with
$\kappa=1.11$, resulting from pulsations of the NS about
equilibrium.  In the bottom
panel, we show the radial acceleration divided by the gravitational
acceleration for the binary system for the
run with $\kappa=1.11$ (solid line) compared the maximum density in
the system (dashed line) as a function of time, finding excellent correlation.}  
\label{fig:rvstkap}
\end{figure}

For an initially synchronized system, we solve this problem by
relaxing the matter in a frame corotating with the binary, and
in the process find the equilibrium value of $\Omega$
self-consistently (See Paper~1, Appendix~A).  
For an irrotational binary, this is not
possible, since we cannot describe a priori 
the proper velocity field toward
which the matter must relax.  We solve the problem instead by a
more direct method.  Realizing that any simple approximation of the
orbital velocity should be near the correct PN value, we alter the
expression for $\Omega$ to include a correction factor which, we hope,
will remove the effect of any small
deviations away from equilibrium on the
gravitational accelerations felt by each NS.  Therefore, we introduce a
parameter $\kappa$ such that
\begin{equation}
\Omega=\kappa \sqrt{\frac{-\dot{v}^{(1)}_x+\dot{v}^{(2)}_x}{2r_0}}.
\label{eq:kap}
\end{equation}
For each value of $\kappa$, we iterate Eqs.~\ref{eq:vx}, \ref{eq:vy},
and \ref{eq:kap} over $\Omega$ to determine a self-consistent
velocity field at $t=0$.  We then perform trials for each value of
$\kappa$, i.e., we perform purely dynamical integrations (with
radiation reaction turned off) for about one full orbit,
until we find one which
produces a nearly circular orbit (such that $r(t)$
changes by no more than $1\%$).  The results of one such set
of calculations, for equal-mass NS with a
$\Gamma=3$ EOS and an initial separation $r_0=4.0$,
is shown in the top panel of Fig.~\ref{fig:rvstkap}.    
We see that for
$\kappa=1.0$, which works extremely well for all runs 
performed in Newtonian gravity, the orbital angular velocity is 
too small, and the pericenter of the orbit lies within the dynamical stability
limit, causing the system to merge (artificially) within the first orbital period.
For $\kappa=1.2$, we see that the orbit is elliptical, with our initial
state corresponding to pericenter.  For a value of $\kappa=1.11$ we obtain
a very nearly circular orbit. A residual small-amplitude
oscillation is shown in greater detail in the middle panel of
Fig.~\ref{fig:rvstkap}.  
We see that it occurs not on the orbital period timescale, but rather on the
internal dynamical timescale of the stars.  It is the direct result of
small-amplitude pulsations of each NS about equilibrium.  This is illustrated
more clearly in the bottom panel of the figure, which shows the clear 
correlation
between the radial acceleration of each NS (shown as a fraction of the
total inward gravitational acceleration) 
and the central density of each NS.  
The numerical noise is an artifact of the small number of particles
located near the very center of each star for the density curve, and of 
precision limits in the calculation for the acceleration curve.  When
averaged over time, we see nearly perfect correlation between the two
quantities.   Also shown in the middle panel of Fig.~\ref{fig:rvstkap}
is a thick solid line depicting the dynamical evolution of a run with
$\kappa=1.11$ and radiation reaction included, to give a sense of the
proper inspiral velocity in relation to the spurious radial velocities
resulting from oscillations.

\subsection{Summary of calculations} \label{sec:summary}

We have performed several large-scale SPH calculations of NS
binary coalescence, assuming an irrotational initial condition for the
binary system.
Table~\ref{table:runs} summarizes the relevant parameters of all runs
performed, listing 
the adiabatic exponent $\Gamma$, the mass ratio
$q$, the initial separation $r_0$, and the number of SPH 
particles.

We continue to use the same nomenclature for our SPH runs
introduced in Paper~2, although we present here a
new run E1, using an improved initial configuration calculated by the
method described in the previous section.
Run E1 is for a system with a
$\Gamma=3$  EOS, equal-mass NS, and an initial separation
$r_0=4.0$, and is similar in all respects to
run B1 of Paper~2 (called the PN run in Paper~1), except that the
initial condition is irrotational.  
It was continued until a quasi-stationary remnant configuration was reached.
Run E2 is for a system with
the same $\Gamma=3$ EOS, but a mass ratio of $q=0.8$, and
was started from a smaller initial separation $r_0=3.5$, since
binaries with smaller masses take longer to coalesce.  
In runs F1 and F2, the EOS has $\Gamma=2$,
but we use the same mass ratios and initial
separations as runs E1 and E2, respectively.  

Additionally, to assess the numerical accuracy and convergence of
calculations for irrotational binaries,
we performed three runs which were primarily designed to test the
dependence of the physical results on numerical parameters.  First, we
performed one run, labeled T2, identical to E1 except for 
a smaller initial separation
$r_0=3.5$.  This run
was used to study how accurately the irrotational flow is maintained
during the early stages of inspiral, and the effects of a small amount of
spurious tidal synchronization on the GW
signal.
We then repeated run T2 using
$5000$ and $500,000$ SPH particles per NS (for a total of $N=10^4$ and
$N=10^6$ particles, respectively), to study the effect of numerical
resolution on calculations where we know small-scale instabilities
will develop.  The number of neighbors was
adjusted in the two runs to be $N_N=50$ and $N_N=200$, respectively. 
This choice is dictated by the convergence and consistency properties
of our basic SPH scheme: convergence toward a physically accurate
solution is expected when {\it both\/} $N\rightarrow\infty$ {\it
and\/} $N_N\rightarrow\infty$, but with $N_N/N\rightarrow 0$ \cite{RasThes}.
The primary consideration behind the choice of the initial separation
at $r_0=3.5$ (rather than $r_0=4.0$) was the high
computational cost of a run with $N=10^6$.

\section{Numerical Results and Tests}
\subsection{Overview}

The coalescence process for binary NS systems is essentially the same
qualitatively whether they are initially synchronized or irrotational.
Prior to merger, both NS show tidal elongation as well as the
development of a tidal lag angle $\theta_{lag}$, as noted in Papers 1
and 2, created as the NS continuously try to maintain equilibrium
while the coalescence timescale gets shorter and shorter.
The inner
edge of each NS rotates forward relative to the binary axis, and the
outer edge of each NS rotates backward.  We define $\theta_{lag}$ to be
the angle in the horizontal plane between the axis of the primary
moment of inertia of each NS and the axis connecting the centers of
mass of the respective NS.  For equal-mass systems, the angle is the
same for both NS.  For binaries with $q<1$ we always find a larger lag
angle for the secondary than the primary. 

\begin{figure}
\centering \leavevmode \epsfxsize=6in \epsfbox{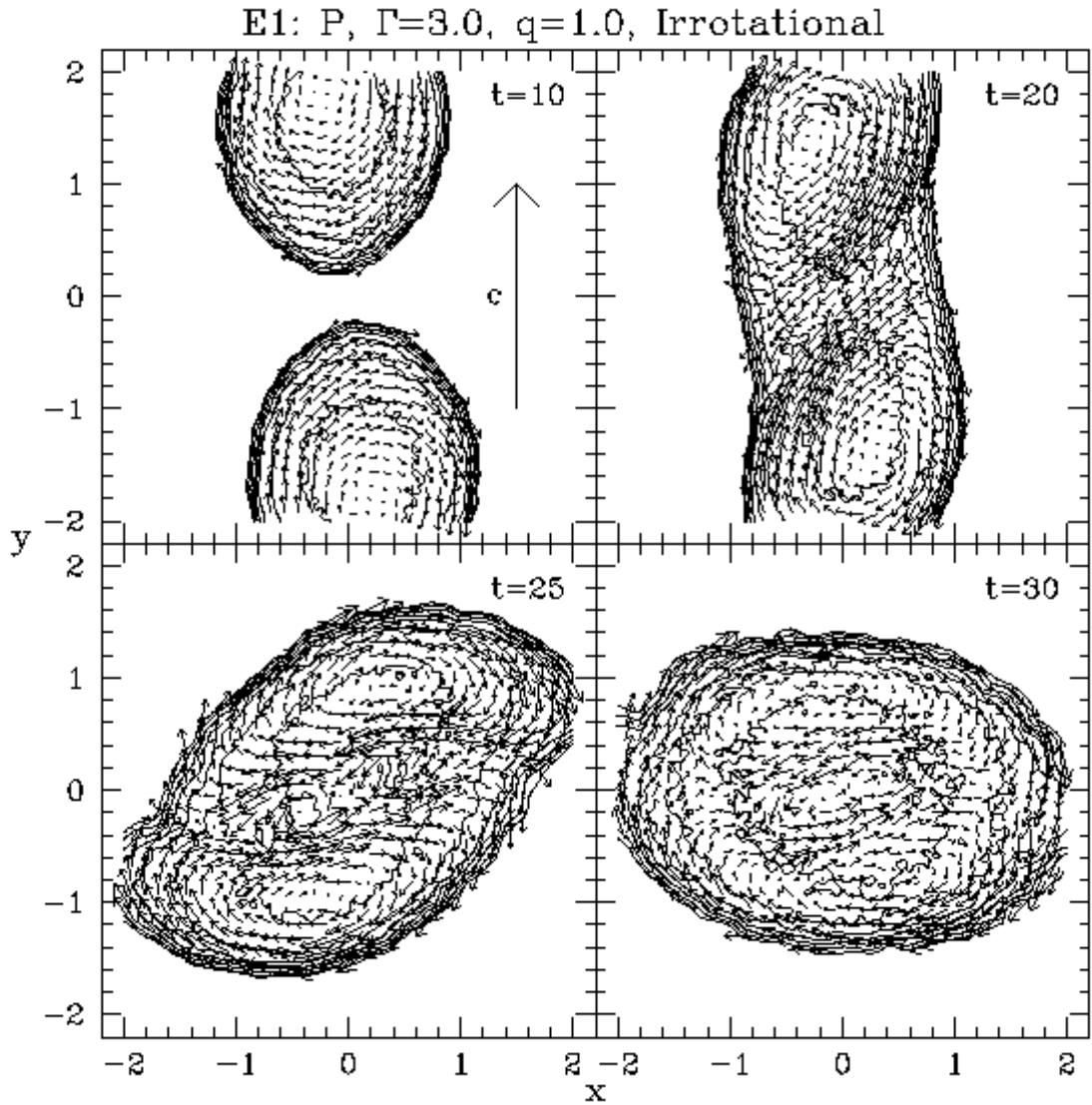}
\caption{Density contours in the orbital plane and velocity field shown in the
corotating frame for the
evolution in run E1 ($\Gamma=3$, q=1).
The physical speed of light, $c_{2.5PN}=2.5$, is
shown to indicate the scale.
Upon first contact of the NS, 
two counterstreaming layers form a turbulent vortex sheet.  As the cores
of the respective NS continue to inspiral, we see the
formation of a merger remnant with the beginnings of
a more coherent differential rotation pattern.}
\label{fig:dvirr}
\end{figure}

Mass shedding through the outer Lagrange
points is suppressed in  irrotational binaries, but the
formation of a differentially rotating remnant is quite similar.
That said, it is important to understand the different computational
challenges presented by  irrotational systems.
To demonstrate this, 
in Fig.~\ref{fig:dvirr} we show the evolution of run E1, with a
$\Gamma=3$ EOS, a mass ratio $q=1.0$, and an initial separation
$r_0=4.0$.  It is in all ways similar to run B1, except that the NS
start from an  irrotational configuration.  Rather than plot
SPH particle positions, we instead show the density contours of the
matter in the orbital plane, 
overlaying the velocity of the material in the corotating
frame, defined by taking a particle-averaged tangential velocity,
such that 
\begin{equation}
\Omega_c=\frac{\sum_i m_i[(xv_y-yv_x)/r_{cyl}]_i}{\sum_im_i(r_{cyl})_i},
\label{eq:omegac}
\end{equation}
where the cylindrical radius is defined as $r_{cyl}=\sqrt{x^2+y^2}$.
In synchronized binaries, the
material maintains only a small velocity in the corotating frame
prior to first contact.  In contrast,
for irrotational binaries material on the inner edge of each NS is
counterspinning in the corotating frame, and thus we see a large
discontinuity in the tangential velocity when first contact is made.
This surface layer, initially at low density, is Kelvin-Helmholtz
unstable to the formation of turbulent vortices on all length scales.
Meanwhile, material on the outer edge of each NS has less angular
momentum in an irrotational binary configuration than in a
synchronized one.  Thus, there is less total angular momentum in the
system, and mass shedding is greatly suppressed, as we will discuss
further in Sec.~\ref{sec:irrotq1}.  At late times, though, the remnant
rotates differentially, with the same profile seen in the synchronized
case, namely $\Omega(r)$ attaining its maximum value at the center of
the remnant, and decreasing as a function of radius.

In Table \ref{table:gw}, we show some basic quantities
pertaining our GW results, as well as the initial time for
all runs, and the tidal lag angle $\theta_{lag}$ which existed at the
moment of first contact.  The negative values of $t_0$ merely reflect the
fact that our runs require more than 20 dynamical times before
reaching peak GW luminosity.  Lag angles $\theta_{lag}$ are
listed for the primary and secondary, respectively, in systems with
$q\ne 1$.  
Each GW signal we compute typically shows an increasing
GW luminosity as the stars approach contact, followed by a
peak and then a decline as the NS merge together.  Most runs then show
a second GW luminosity peak of smaller amplitude.  For all
of our runs we list the maximum GW luminosity $L_{max}\equiv {\rm
max}(L_{GW}(t)$ and
maximum GW amplitude $h_{max}\equiv {\rm
max}(\sqrt{h_+(t)^2+h_{\times}(t)^2})$, where $h_+$, $h_{\times}$, and
$L_{GW}$ are defined by Eqs.~(23)-(25) of Paper~1 (see also
Eqs.~\ref{eq:hpl} and \ref{eq:hcr} below).
Quantities referring to the first and second luminosity peaks are
denoted ``1'' and ``2'', respectively.  In addition, we
show the time $t^{(2)}$ at which the second peak occurs.

We continued several of our runs to late times to study the full
GW signal produced during the coalescence, as well as to
study the properties of the merger remnants that may form in these
situations.  For each of these runs, we list several of the basic
parameters of the merger remnant in Table \ref{table:final}, using the
values computed for the remnant at $t=65$.  We
identify the remnant mass $M_r$, defining the edge of the remnant by
a density cut $r_*>0.005$, as well as the gravitational mass of the
remnant in the PN runs, where the gravitational mass, which differs
from the rest mass, is given by $M_{gr}\equiv\int r_*(1+\delta)d^3x$
(see Paper~1 for more details).
Additionally, we list the Kerr parameter $a_r\equiv cJ_r/M_{gr}^2$, 
central and equatorial values of the angular
velocity, $\Omega_c$ and $\Omega_{eq}$, the semi-major axis $a_1$ and
ratios of the equatorial and vertical radii $a_2/a_1$ and $a_3/a_1$,
and the ratio of the principal moments of inertia $I_2/I_1$.

\subsection{Dependence on EOS Stiffness}
\label{sec:irrotq1}

\begin{figure}
\centering \leavevmode \epsfxsize=6in \epsfbox{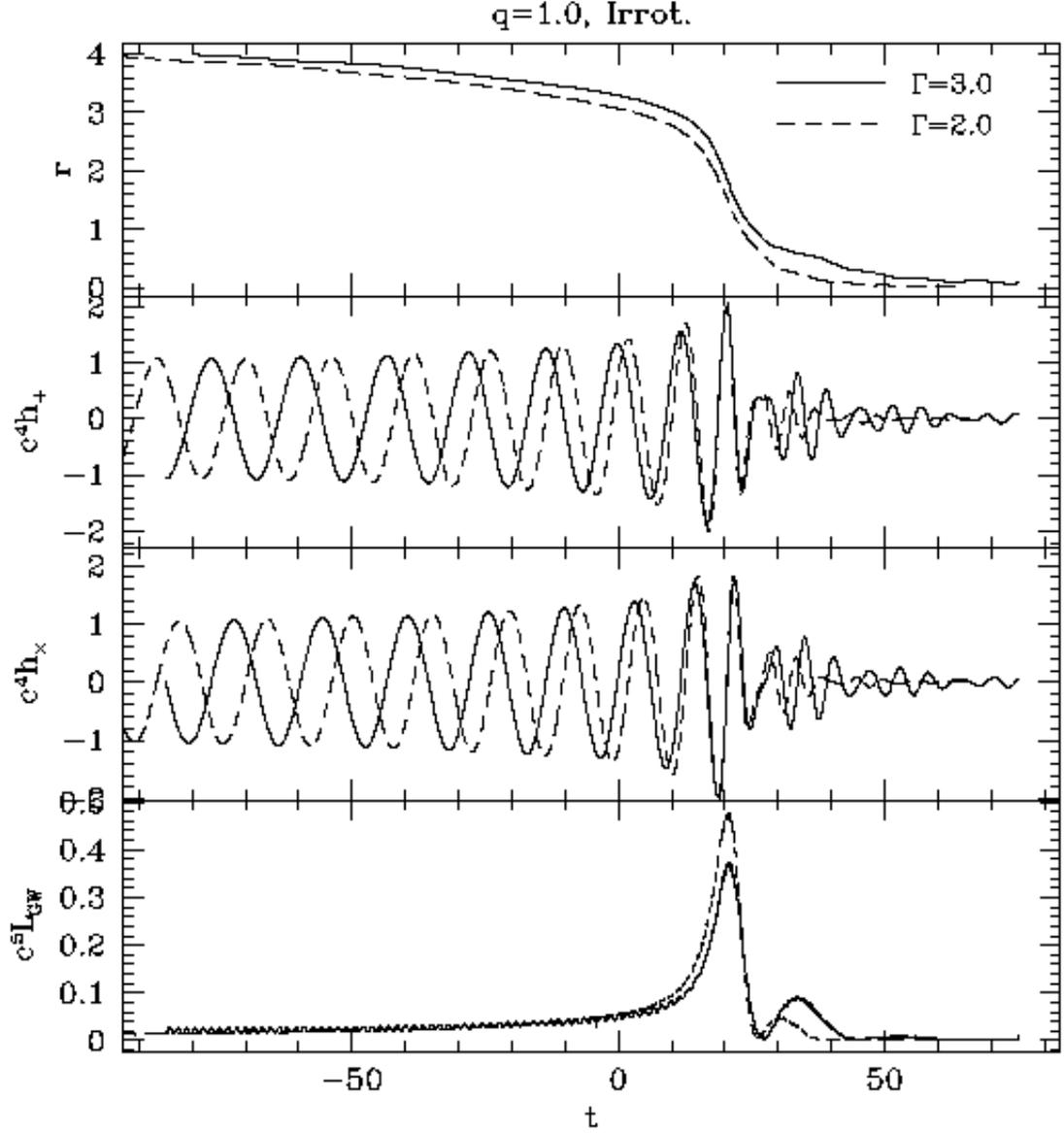}
\caption{Binary separation (top panel), GW amplitudes in both
polarizations (middle
panels) and GW luminosity (bottom panel) as a function of
time for runs E1 and F1.
The solid lines correspond to run E1 ($\Gamma=3$), the dashed lines to
run F2 ($\Gamma=2$).  The dynamical stability limit for the softer EOS lies
within that of the stiffer one.  The softer EOS also results in a
higher peak GW luminosity, but smaller amplitude post-merger GW
emission.}
\label{fig:gweos}
\end{figure}

To study the effect of the EOS on the evolution of
irrotational NS binaries, we calculated mergers of both
$\Gamma=3$ and $\Gamma=2$ polytropes (runs E1 and F1, respectively).  
A comparison of the binary separations as a function of time, along with the 
dimensionless GW luminosities and amplitudes
are shown in Fig.~\ref{fig:gweos}.
Immediately apparent is a difference in the location of the dynamical
stability limit found in the two calculations.
The orbit of the binary system containing NS with a 
softer EOS (run F1) remains stable at separations where the binary system
with the stiffer EOS has
already begun to plunge inward dynamically.
We see that, as in the
synchronized case presented in Paper~2, the peak GW luminosity in
dimensionless form is
larger for the softer EOS, but after a secondary luminosity
peak the remnant relaxes toward a spheroidal, non-radiating
configuration (with essentially no emission whatsoever after $t\simeq
40$).  We also note that the secondary peak occurs sooner after the
primary peak, by a factor of $\simeq 30\%$.

\begin{figure}
\centering \leavevmode \epsfxsize=6in \epsfbox{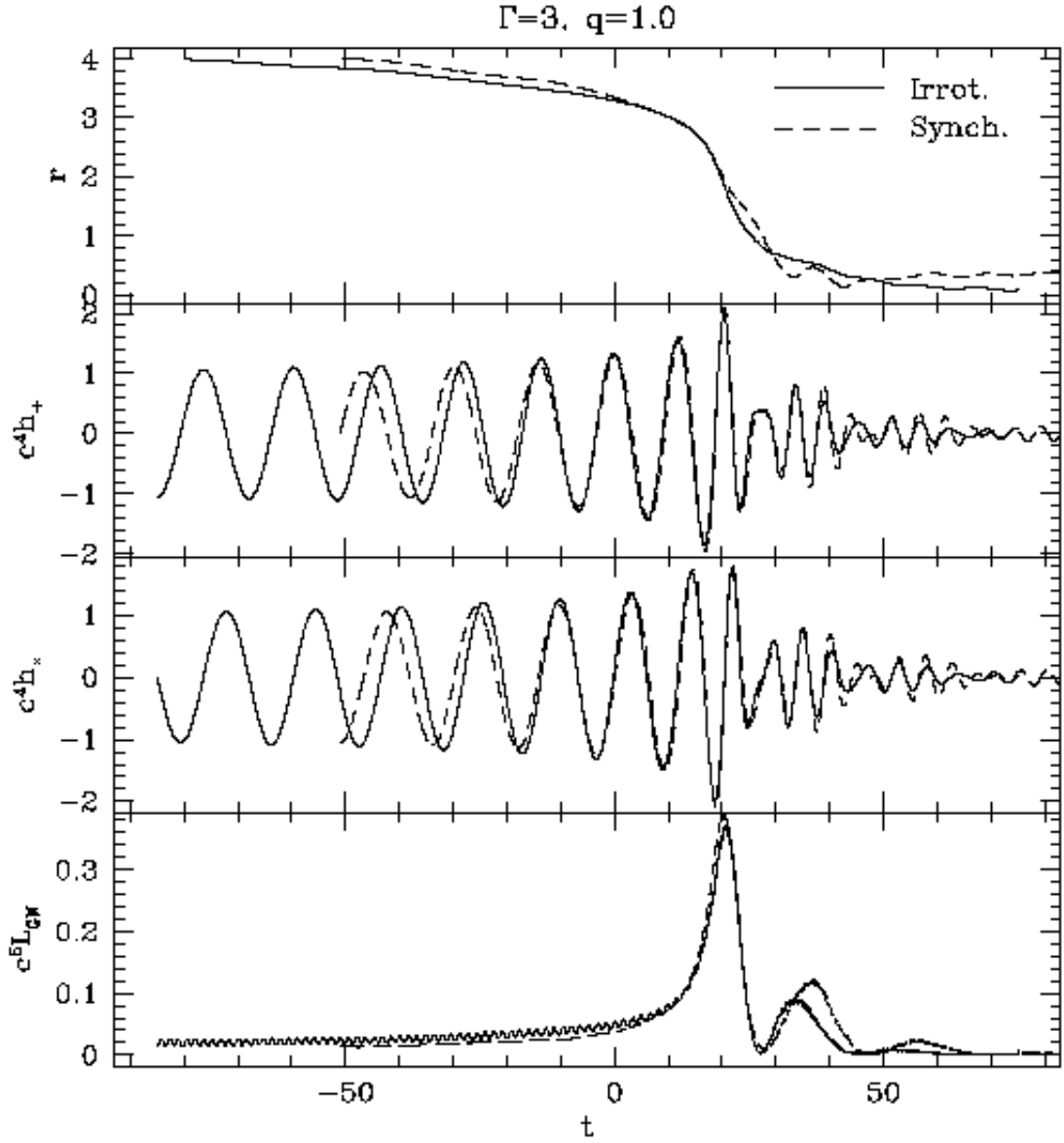}
\caption{Binary separation (top panel), GW amplitudes (middle
panels) and GW luminosity (bottom panel) as a function of
time for runs E1 and B1, started from an
irrotational (solid line) and a synchronized (dashed line) initial condition,
respectively.
The synchronized run contains more energy and is relatively more
dynamically unstable.  While the initial peaks in the GW
luminosities are of similar amplitude, the secondary peaks are much
more luminous for the synchronized binary.}
\label{fig:gwspin}
\end{figure}

Even though they are presumed to be unphysical, calculations started
from a synchronized initial condition make up much of the body of work
performed to date on the binary NS coalescence problem.  Noting this,
we compare our irrotational run E1, with parameters detailed above,
to one similar in every respect but
started with a synchronized initial condition, our run B1.  In the top panel of
Fig.~\ref{fig:gwspin}, we see that the inspiral tracks do not align
particularly well.  Synchronized binaries contain more total energy,
and are thus less dynamically stable than irrotational ones, leading to a
more rapid inspiral, even before the stability limit is reached.
Additionally, the binary separation ``hangs up'' earlier, at a
separation $r\approx 2.0$, indicating the onset of mass shedding,
as matter begins to expand radially outward from the system.
The middle and bottom panels of the figure show the GW signals
and luminosity, respectively, for the two runs.  While the initial
luminosity peaks are similar for 
both runs, both in amplitude and morphology, the secondary peaks are
vastly different.  The secondary peak for the synchronized system
is of considerably greater magnitude than in the irrotational case, and delayed
relative to it.  We conclude that, while the adiabatic index seems to be the
dominant factor in determining the GW signal during the merger
itself, the initial velocity profiles 
of the NS play a key role in the evolution
of the remnant, as well as affecting the orbital dynamics during inspiral.

\begin{figure}
\centering \leavevmode \epsfxsize=6in \epsfbox{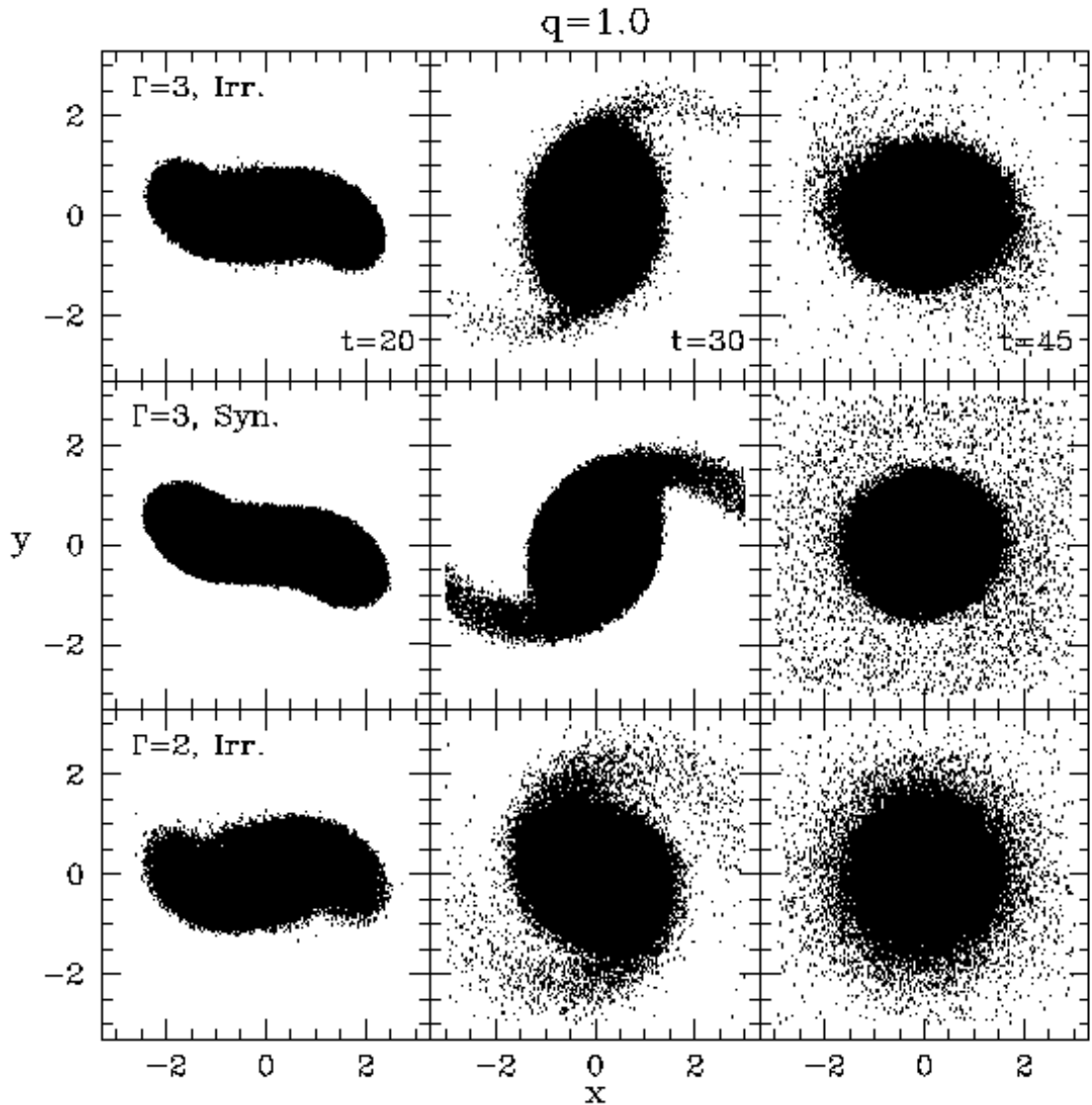}
\caption{Particle plots for runs E1 ($\Gamma=3$; top panels), 
B1 (synchronized, $\Gamma=3$; middle panels), and F1 ($\Gamma=2$;
bottom panels), described in
Figs.~\protect\ref{fig:gweos} and \protect\ref{fig:gwspin}.
The orbital rotation is in the counter-clockwise direction.
The plots show projections of all SPH particles at $t=20$ (left),
$t=30$ (center), and $t=45$ (right).  Mass shedding is
more sensitively dependent on the initial velocity field, but the remnant
equatorial ellipticity, and thus the post-merger GW emission, is
dominated by the choice of EOS.}
\label{fig:xyeos}
\end{figure}

To better understand the features found in the GW signals of
these calculations, particle plots for runs E1, B1, and F1
are shown in Fig.~\ref{fig:xyeos}.  Comparing the leftmost
panels, we see that at $t=20$, when the GW
luminosity peaks, the mass configurations are qualitatively
similar, although more low-density material is seen at the edges of the newly
forming remnant in the run with the $\Gamma=2$ EOS, 
conforming to the general density profile expected for a softer EOS.  
More subtle is the greater extension of the matter
found in the synchronized run.
Since material on the outer edge of each NS has greater angular momentum
when a synchronized initial condition is chosen, the calculation shows
that such an initial condition allows for greater efficiency at channeling 
material outward during the
final moments of inspiral.  This difference is made abundantly clear
by a comparison of the calculations at $t=30$, shown in the center
panels.  We see extensive mass
shedding from the synchronized run, much less from the irrotational
runs.  There is significantly more mass shedding from the run with
the softer EOS, but most of the material remains extremely close to
the remnant.  Finally, by $t=45$, we see that the softer EOS produces
a nearly spherical remnant, whereas the calculations with a stiffer
choice of EOS produce remnants which are clearly ellipsoidal, and will
continue to radiate GWs for some time, albeit at a much
lower amplitude than at the peak.

\begin{figure}
\centering \leavevmode \epsfxsize=6in \epsfbox{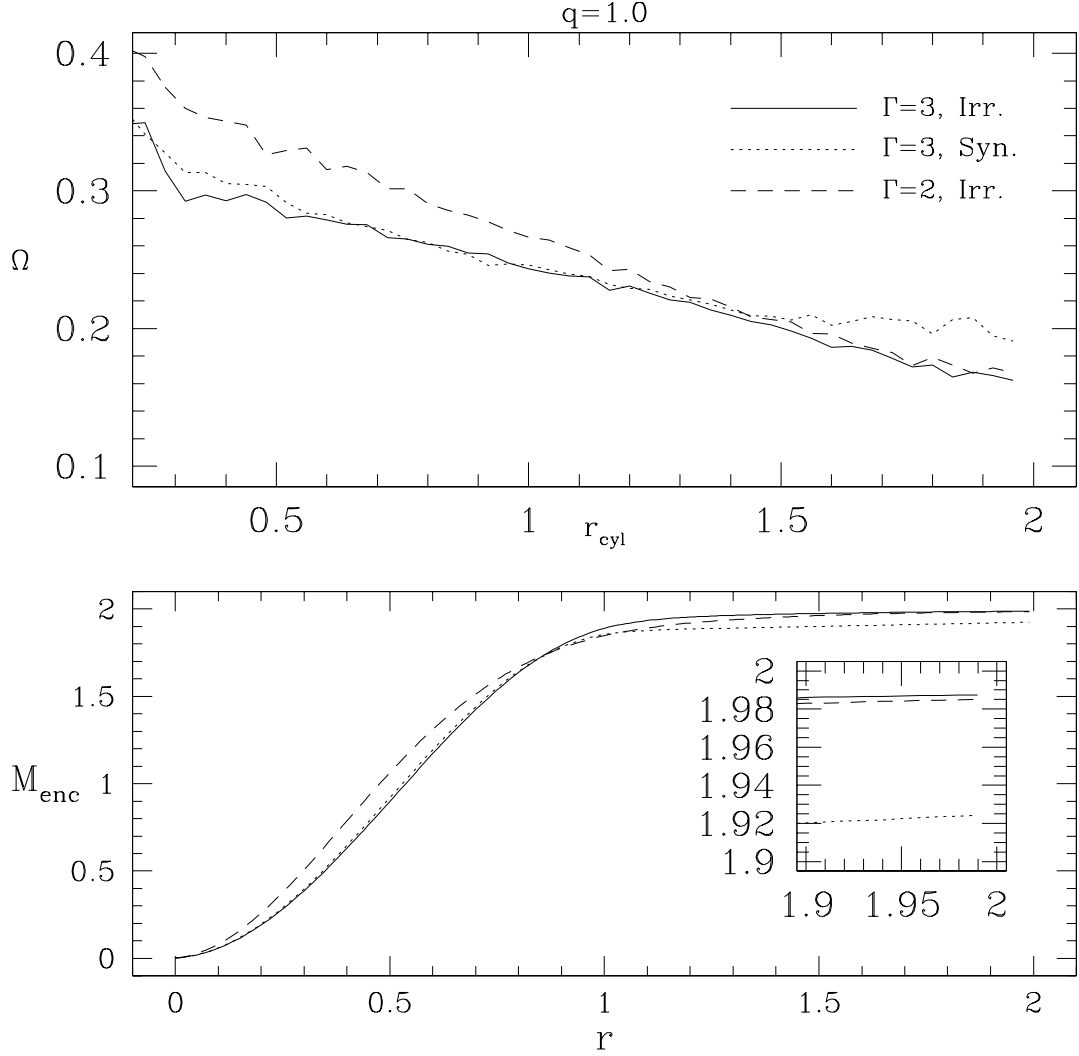}
\caption{Angular velocity as a function of cylindrical radius (top
panel) and enclosed mass as a function of radius 
for the remnants of runs B1, E1, and F1, described in
Figs.~\protect\ref{fig:gweos} and \protect\ref{fig:gwspin}.   
The profiles, all taken at $t=65$, correspond to irrotational run E1 
($\Gamma=3$; solid line), run F1 ($\Gamma=2$; dashed line), and 
run B1 (synchronized, $\Gamma=3$; dotted line).}
\label{fig:finaleos}
\end{figure}

The strong influence of the choice of both EOS and initial velocity
profile on the final
state of the remnant is shown in Fig.~\ref{fig:finaleos} for the same
three runs.  In the top
panel, we see the angular velocity profiles of the remnants at $t=65$.
We see that the choice of EOS plays an important role near the center
of the remnant, but at $r>1.0 $, the velocity profiles are
essentially identical.  The pattern holds as well for the mass
profiles, which are shown in the bottom panel.  The softer EOS leads
to a more centrally condensed remnant, as we would expect, but 
the remnants formed in both irrotational calculations
contain virtually all the system mass within $r\approx 2.0$, with no
more than $1\%$ escaping to larger radii.  There is slightly more mass
shedding past $r>2.0$ for the softer choice of EOS, since more of
the low-density material originally found at the edges of the NS is
shed through the outer Lagrange points of the system.  By comparison,
the  synchronized run sheds almost $5\%$ of the total system
mass past $r>2.0$, even though the angular velocity profile at small
radii is nearly the same as for the irrotational run with the same
choice of EOS.

\subsection{Unequal-mass Binaries}
\label{sec:q8}

\begin{figure}
\centering \leavevmode \epsfxsize=6in \epsfbox{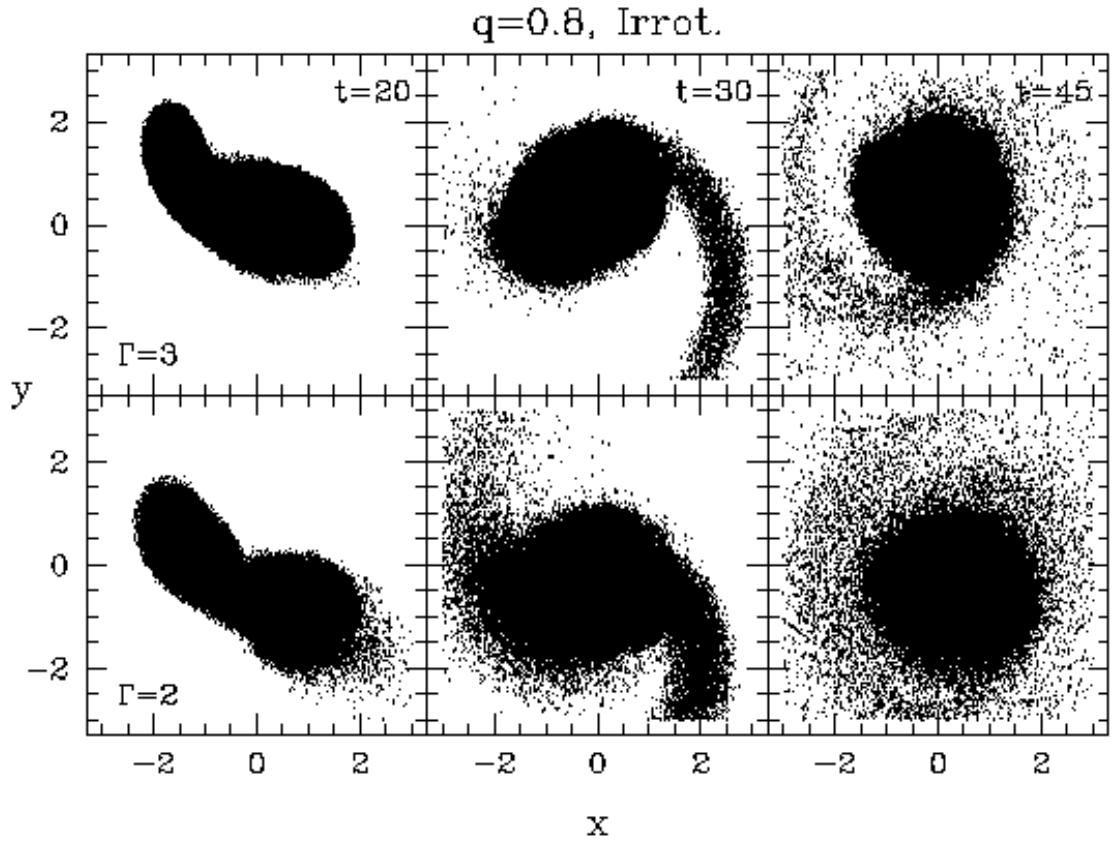}
\caption{Particle plots for runs E2 ($q=0.8$, $\Gamma=3$; top panels)
and F2 ($q=0.8$, $\Gamma=2$; bottom panels), respectively.
From left to right, we see
snapshots taken at $t=20$, $t=30$, and $t=45$, showing in both cases
the tidal disruption of the secondary, the formation of a single
spiral arm during mass shedding, and the eventual creation of a
massive torus around the merger remnant.}
\label{fig:xyq8}
\end{figure}

Even though all well-measured NS masses in relativistic binary pulsars
appear roughly consistent with a single NS mass $M_{NS}\approx 1.4
M_{\odot}$ \cite{TC}, it is important to consider cases where the
two NS have somewhat different masses.  The roles played by the primary and
secondary in unequal-mass binary mergers are remarkably 
different compared to the picture
developed above for equal-mass systems.  In Paper~2, we found that
the primary NS in the system
remains virtually undisturbed, simply settling at the center
of the newly formed merger
remnant.  The secondary is tidally disrupted prior to merger, forming
a single thick spiral arm.  Most of the material originally located in
the secondary eventually forms the outer region of the merger remnant,
but a significant amount of material is shed to form a thick torus
around the central core.  In Fig.~\ref{fig:xyq8}, we show particle plots
for runs E2 and F2, with a $\Gamma=3$ and $\Gamma=2$
EOS, respectively, both with mass ratio $q=0.8$.
In the leftmost panels, at $t=20$, we see that the secondary, located on the
left, is tidally disrupted as it falls onto the primary.  For the
softer, $\Gamma=2$ EOS (run F2), we see a greater extension of
the secondary immediately prior to merger, as well as early mass
shedding from the surface of the primary, as material from the
secondary essentially blows it off the surface of the newly forming
remnant.  This process continues, so that by $t=30$ (center panels),
the secondary has begun to shed a considerable amount of mass in a
single spiral arm which wraps around the system.  Much like in the
equal-mass case, the spiral arm is much broader for the softer EOS.
Mass loss from the primary is greatly reduced in the system with the
stiffer EOS, with only a scattering of particles originally located in
the primary lifted off the surface.  Finally, by $t=45$ (right
panels), we see that the spiral arm has in both cases begun to
dissipate, leaving a torus around the merger remnant containing
approximately $3-4\%$ of the total system mass.

\begin{figure}
\centering \leavevmode \epsfxsize=6in \epsfbox{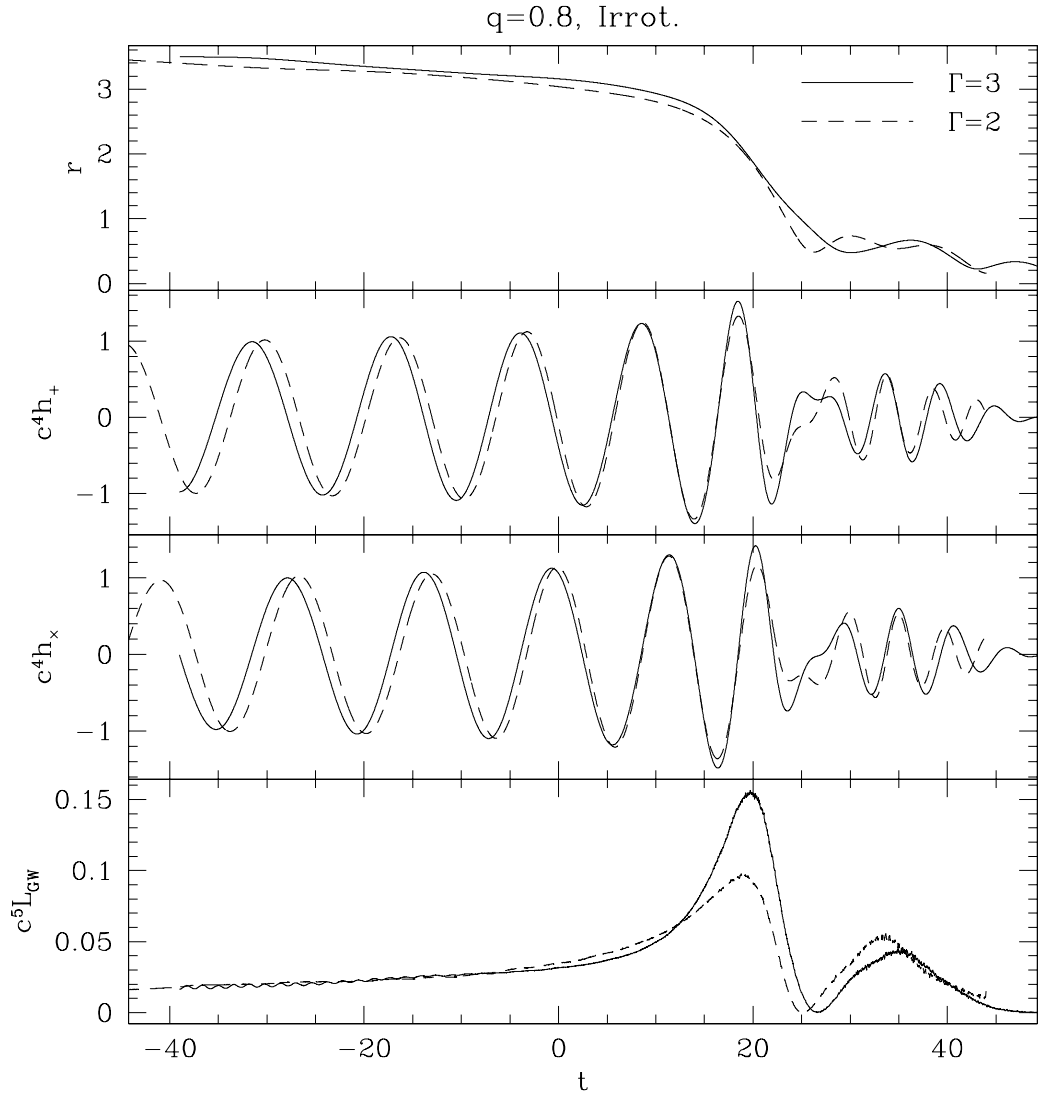}
\caption{Binary separation (top panel), GW amplitudes (middle
panels) and GW luminosity (bottom panel) as a function of
time for our irrotational runs E2 and F2, 
as shown in Fig.~\protect\ref{fig:xyq8}.  The
solid and dashed curves correspond to run E2 ($\Gamma=3$) and run
F2 ($\Gamma=2$), respectively.  We see that GW production
is significantly suppressed for the softer $\Gamma=2$ EOS during the
first peak.}
\label{fig:gwq8}
\end{figure}

Although the merger process is significantly different for equal-mass
and unequal-mass binaries, the details of the inspiral phase are
reasonably similar.  In particular, the evolution of the binary
separation for runs with $q=0.8$, shown in the top panel of
Fig.~\ref{fig:gwq8} is roughly similar to what was found in
Fig.~\ref{fig:gweos} for binaries with $q=1.0$.  For both choices of
the EOS, the dynamical stability limit is located at approximately
the same separation for both $q=0.8$ and $q=1.0$ binaries, although
prior to the onset of instability the more massive equal-mass binaries
show a more rapid stable inspiral.  As we found before, the dynamical
stability limit occurs further inward for the softer choice of the NS
EOS.

In Paper~2, we found that the scaling of the maximum GW amplitude
and luminosity as a function of the system mass ratio 
followed a steeper power law in synchronized binaries than would be
predicted by Newtonian point-mass estimates.  
Newtonian physics predicts that $h_{max}\propto q$ and $L_{max}\propto
q^2(1+q)$ for merging binary systems.  
RS found the scaling in their numerical
calculations to roughly follow empirical power law relationships given
by $h_{max}\propto q^2$ and $L_{max}\propto q^6$.  The discrepancy
results from the unequal role played by the two components during the
final moments before plunge.  The primary, which remains relatively
undisturbed, contributes rather little to the GW signal,
especially during the final moments before coalescence.  Thus, the
GW power is reduced as the mass ratio is decreased.  
Similar results were found in Paper~2 for PN calculations of
synchronized binaries, especially a steeper decrease in the GW
luminosity as a function of the mass ratio for binaries with a
soft EOS.

In the middle and bottom panels of Fig.~\ref{fig:gwq8} we show the
GW signals and luminosity, respectively, for 
runs~E2 and~F2.  We find the same strong decrease in the GW
power, especially for the softer $\Gamma=2$ EOS.  We expect that there
should be a strong observational bias towards detecting mergers of equal-mass
components, especially if the EOS is softer.  

\subsection{Dependence on Initial Separation}
\label{sec:sep0}

\begin{figure}
\centering \leavevmode \epsfxsize=6in \epsfbox{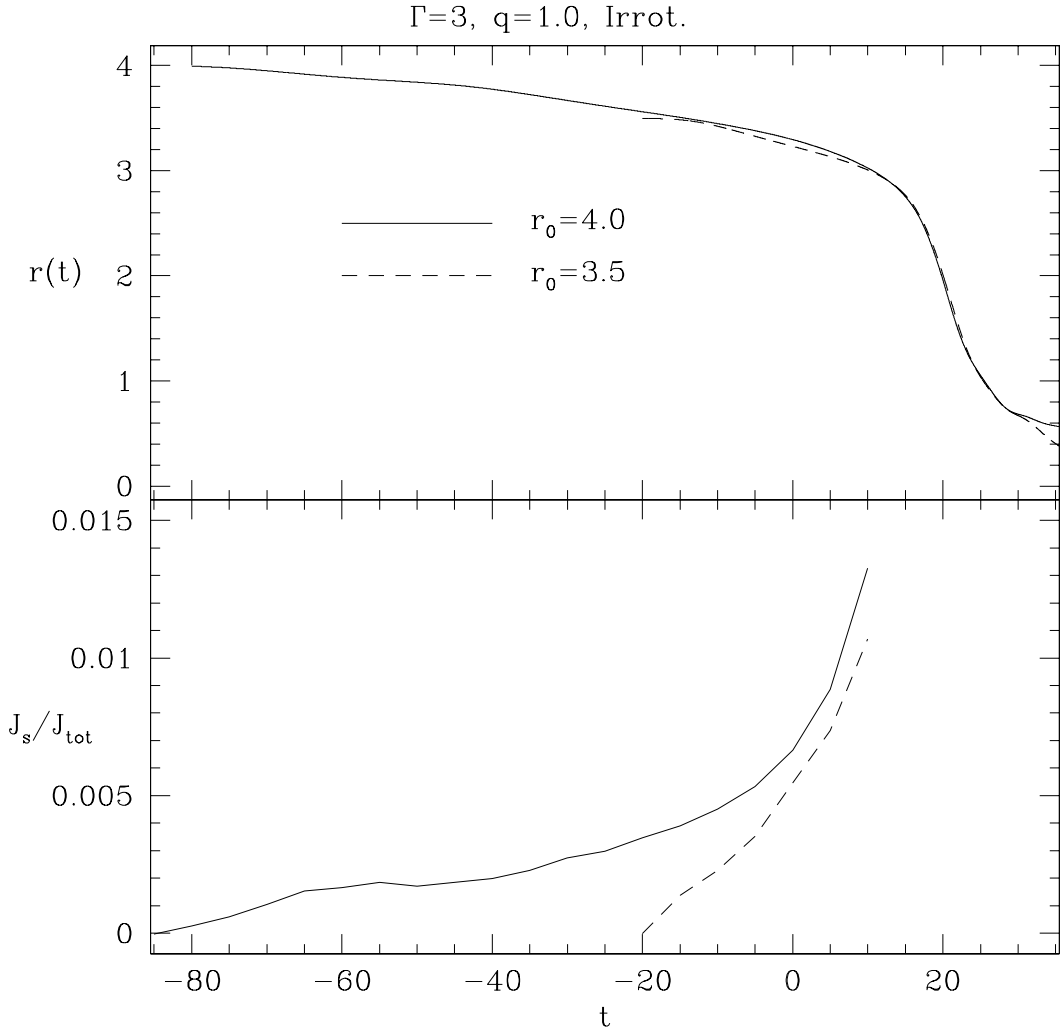}
\caption{Binary separation (top panel) and the 
ratio of each star's spin angular momentum to its total angular momentum
(bottom panel) as a function of time for runs E1 and T2 which
started from an initial separation
$r_0=4.0$ (solid line) and $r_0=3.5$ (dashed line), respectively.  
The binary
with larger initial separation develops greater spin angular
momentum throughout the calculation prior to merger.}
\label{fig:rvstsep0}
\end{figure}

All our results presented above for equal-mass binaries used an
initial binary separation $r_0=4.0$, in part for the sake of
comparison with previous calculations for synchronized binaries. 
This choice agrees with the standard approach of utilizing the
largest possible separation for which the calculation can be performed
using a reasonable amount of computational resources.  This also has
the advantage that any small deviations from equilibrium will
generally be damped away before the NS actually make contact.  It also
allows for the best determination of the dynamical stability limit,
However, a large initial separation can create problems
in the case of
irrotational binaries, since numerical shear viscosity inherently
present in SPH codes can lead to some degree of tidal synchronization
of the NS during the inspiral phase.  Thus, by the time the merger
takes place, the NS will no longer be completely irrotational.  To
study this effect, we calculated mergers for equal-mass 
NS with a $\Gamma=3$ EOS starting at initial separations of both
$r_0=4.0$ (the aforementioned run E1) and $r_0=3.5$ (run T2).  
In the top panel of Fig.~\ref{fig:rvstsep0}, we show the
binary separation as a function of time for both runs, noting the good
agreement throughout.  At the very end, during the merger
itself, we do
see the beginning of a slight discrepancy, attributable in large part to
greater mass shedding in the calculation started at greater
separation.
This is similar to what was seen in Sec.~\ref{sec:irrotq1}, where
run B1, which had greater spin angular momentum, showed
greater mass shedding, but the effect is greatly reduced in magnitude
here since the NS in run T2 are nowhere near complete synchronization at the
moment of first contact.

In the bottom panel of the figure, we plot
the ratio of the net spin angular momentum of the NS about
their own centers of mass to the total angular momentum of the binary
system, as a function of time.  We see that the NS do
gradually acquire a rotation pattern which corresponds to the
direction of corotation, although there is nowhere near enough time to  
synchronize the binary.  The effect is greatly enhanced immediately prior to
merger in both cases, as the NS develop tidal lag angles and become
distorted.  
By the time the binary initially started from $r_0=4.0$ 
reaches a separation
of $r=3.5 $, the net angular momentum around each NS around its
center of mass
is equal to approximately $0.5\%$ the value we would expect should the
binary be synchronized.  This difference persists throughout the
inspiral phase when the two calculations are compared. 
  
\begin{figure}
\centering \leavevmode \epsfxsize=6in \epsfbox{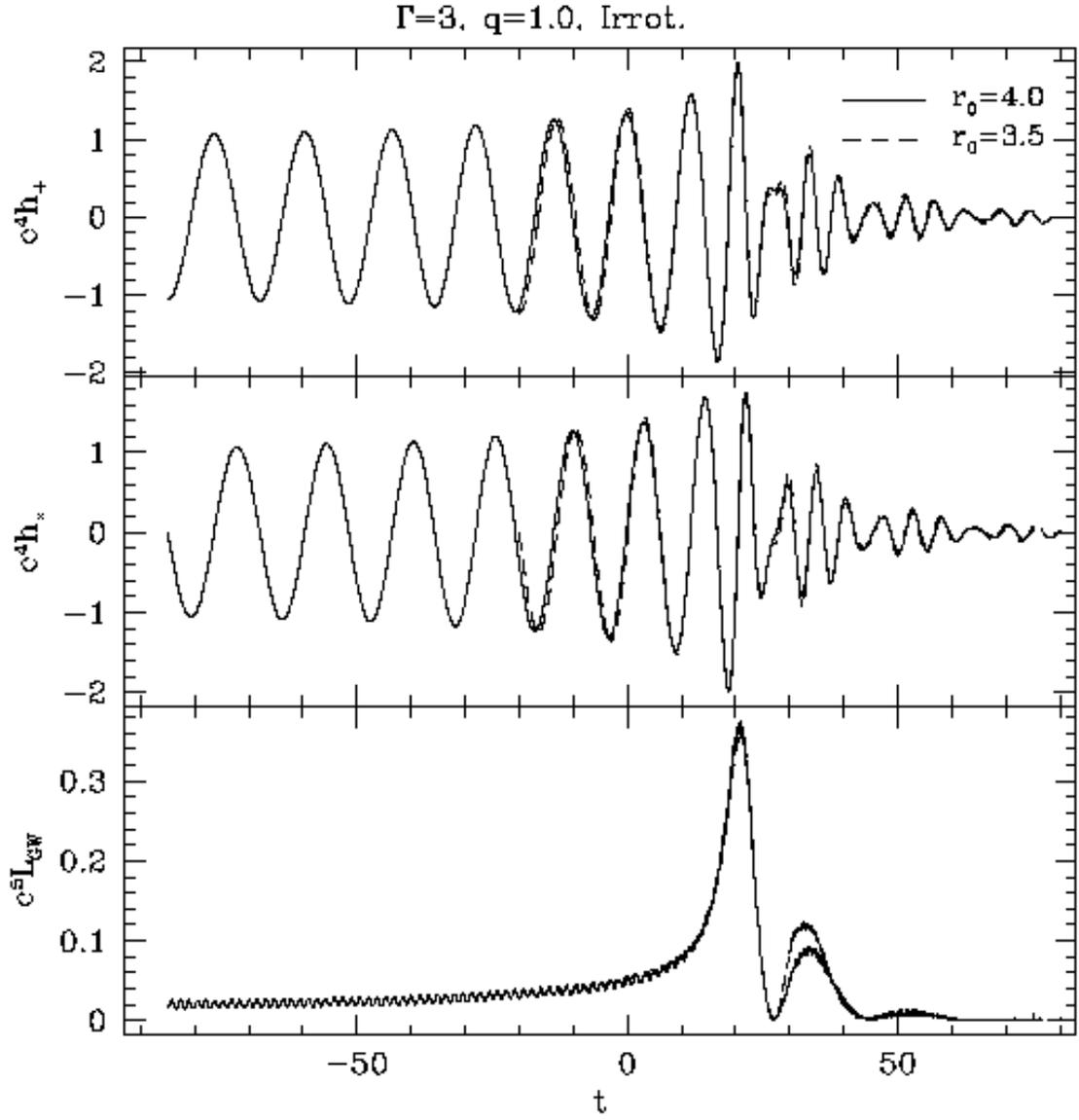}
\caption{GW amplitudes in both polarizations 
(top and middle panel) and GW luminosities (bottom panel) 
for the two runs described
in Fig.~\protect\ref{fig:rvstsep0}.  The agreement is
excellent throughout, except for a slight difference in the amplitude
of the second GW peak.}
\label{fig:gwsep0}
\end{figure}

In Fig.~\ref{fig:gwsep0}, we compare the GW signals and luminosities
for the two runs.  We find excellent agreement between the two waveforms, 
both in amplitude and in phase.  Both runs show the modulated, damped
GW luminosity which is characteristic of all runs we have
computed using PN gravity.
There is a
slight difference in the amplitude of the signal during the second
GW luminosity peak, but we expect such differences to be
minor in light of such issues as the uncertainty in the equation of
state and the larger problem of a proper relativistic treatment of
gravitation.  

\begin{figure}
\centering \leavevmode \epsfxsize=6in \epsfbox{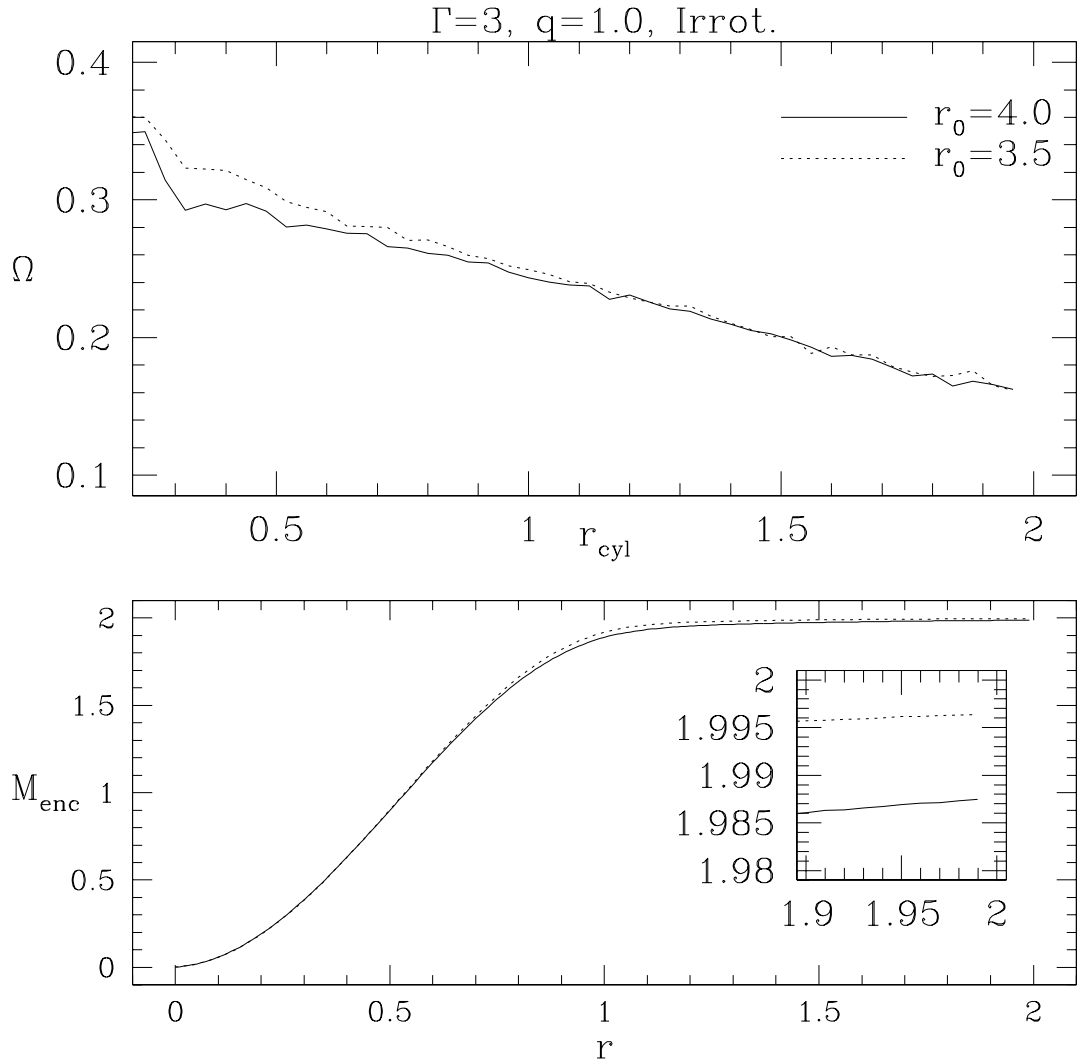}
\caption{Angular velocity as a function of cylindrical radius (top
panel) and enclosed mass as a function of radius
for the remnants of the
runs shown in Fig.~\protect\ref{fig:rvstsep0}.  The profiles are taken
at $t=65$ for both runs.  We see that mass shedding to larger radii is
very slightly increased for the calculation started from $r_0=4.0$, but that
in both cases virtually all the matter in the system ends up in the
remnant itself.  The inset shows the profile at the outer edge of the
system, indicating that no more than $\simeq 0.5\%$ of the material is
ejected to larger radii.}
\label{fig:finsep0}
\end{figure}

To focus on the effect of varying $r_0$ on the 
final results, we show the final mass and angular velocity
profiles of the remnants for the two calculations in
Fig.~\ref{fig:finsep0}.  The results are in good agreement, although we
see that the greater spin angular momentum of the run started at greater
initial separation leads to approximately three times as much 
 mass being deposited in a
halo which surrounds the remnant while remaining gravitationally bound
to it.  In both cases, however, the total mass in the halo is less
than $1\%$ of the total system mass.
The inner region of the remnant in the run started from $r_0=4.0$ actually
spins slightly slower than in the run started further inward, even though
the NS have a greater spin angular momentum at the moment of first
contact, but only because
angular momentum transport outward was marginally more efficient in
this case.  We conclude that the choice of initial separation plays
very little role in determining the results of our
calculations, so long as we start from an initial separation 
$r_0\gtrsim 3.5$. 

\subsection{Dependence on Numerical Resolution}

As discussed in Sec.~\ref{sec:summary}, 3D calculations with limited
spatial resolution could lead to GW signals which are dependent upon
the number of particles used.
To test this, 
GW signals computed from NS merger calculations, we
performed runs T1, T2, and T3, which all have equal-mass NS, start
from the same initial separation $r_0=3.5$, and use a $\Gamma=3$ EOS,
but vary by two orders of magnitude in the number of SPH particles
used, from run T1 with $10^4$ to run T3 with $10^6$ SPH particles.  
Although we could have used $r_0=4.0$ as an initial separation, we
felt that the smaller initial separation was justified given the large
computational overhead required to do a calculation with a million SPH
particles.  To the best of our knowledge, Run T3 
is the largest and highest resolution SPH calculation of 
irrotational binary NS coalescence to date.

\begin{figure}
\centering \leavevmode \epsfxsize=6in \epsfbox{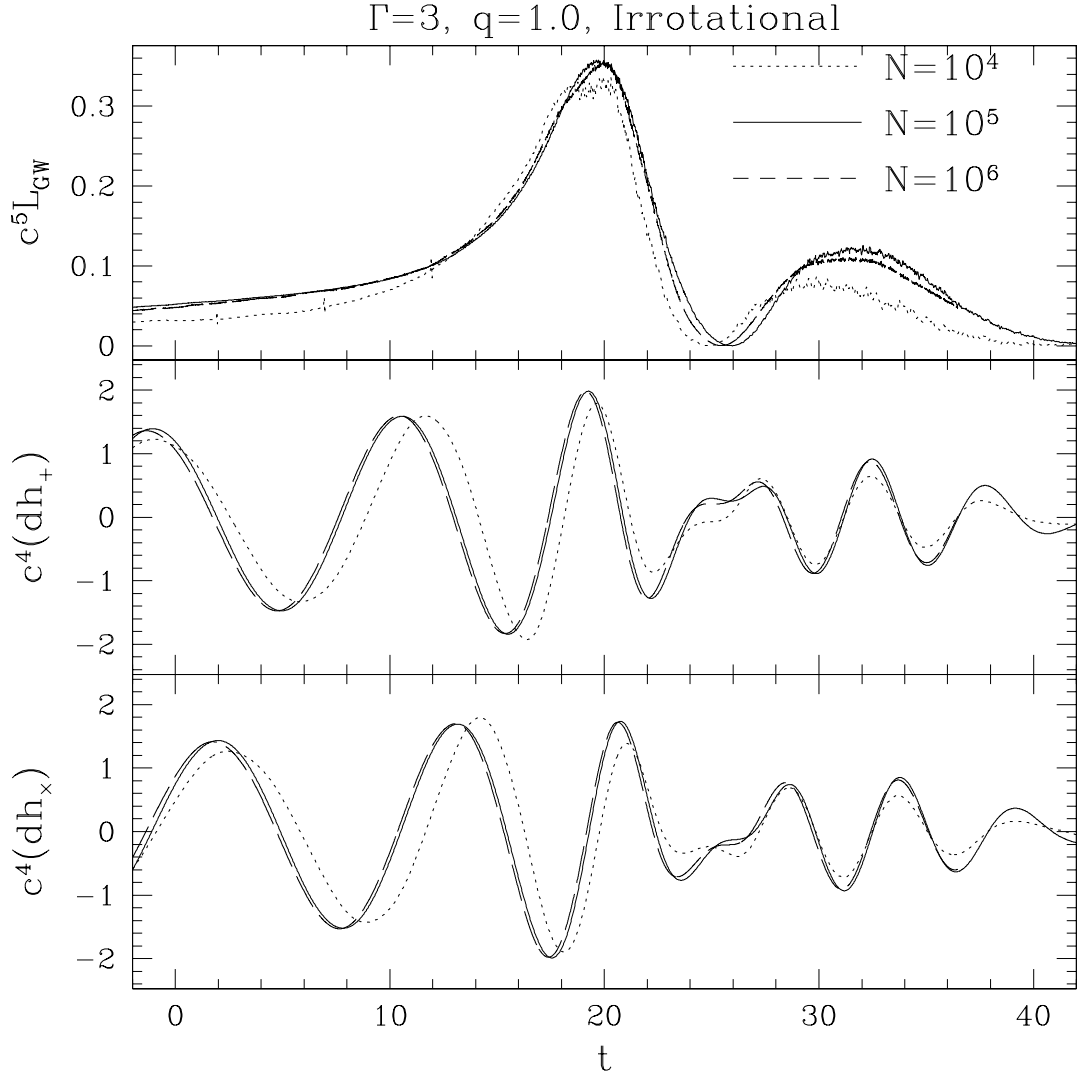}
\caption{GW luminosities (top) and waveforms (middle and
bottom) for calculations using
increasing number of particles.  The dotted lines correspond to run T1
($N=10^4$), the solid lines to run T2 ($N=10^5$), and the dashed lines to
run T3 ($N=10^6$).  We see that the two highest resolution runs agree almost
perfectly.  The lowest resolution run is more susceptible to initial
deviations from equilibrium, and shows some significant differences
from the higher resolution runs, especially after the first GW
luminosity peak.}
\label{fig:gwnp}
\end{figure}

A comparison of the GW signals in both polarizations, as
well as the GW luminosities, is shown in Fig.~\ref{fig:gwnp}.
We see that the
lowest resolution run T1 
produces a GW signal clearly different than higher resolution 
runs T2 and T3.  The difference is due in part to initial
oscillations of the NS about quasi-equilibrium.  Such oscillations are
greatly reduced by increasing the number of SPH particles.
Since all three calculations
started with the same approximate ellipsoidal
models (see Sec.~\ref{sec:initconf}),
we conclude that the amplitudes of the initial
fluctuations result primarily from
numerical noise.  The two runs
with higher resolution show good agreement from beginning to end,
producing nearly identical GW luminosities.  Reassuringly, the
GW signals also remain in phase
throughout, indicating that calculations with $N\gtrsim 10^5$ can
indeed model well these irrotational binary mergers.
To estimate the precision which can be achieved for these
numerical resolutions, we show in
Table~\ref{table:gravnp} a comparison of GW quantities
computed at various times for each of the three runs.  At $t=10$,
$t=20$, and $t=30$, we show the total GW strain $h(t)$,
and the instantaneous frequency of the GWs
$\Omega_{GW}(t)\equiv\dot{\theta}_{GW}(t)$, defined by the relations
\begin{eqnarray}
h_+(t)&\equiv&h(t)\cos\theta_{GW}(t), \label{eq:hpl}\\
h_{\times}(t)&\equiv&h(t)\sin\theta_{GW}(t). \label{eq:hcr}
\end{eqnarray}
We see that for the two high resolution runs, no
quantity shown varies by more than about $2\%$, whereas the difference
is more than $10\%$ in the computed GW strains
at late times between our highest and lowest resolution runs.

\begin{figure}
\centering \leavevmode \epsfxsize=6in \epsfbox{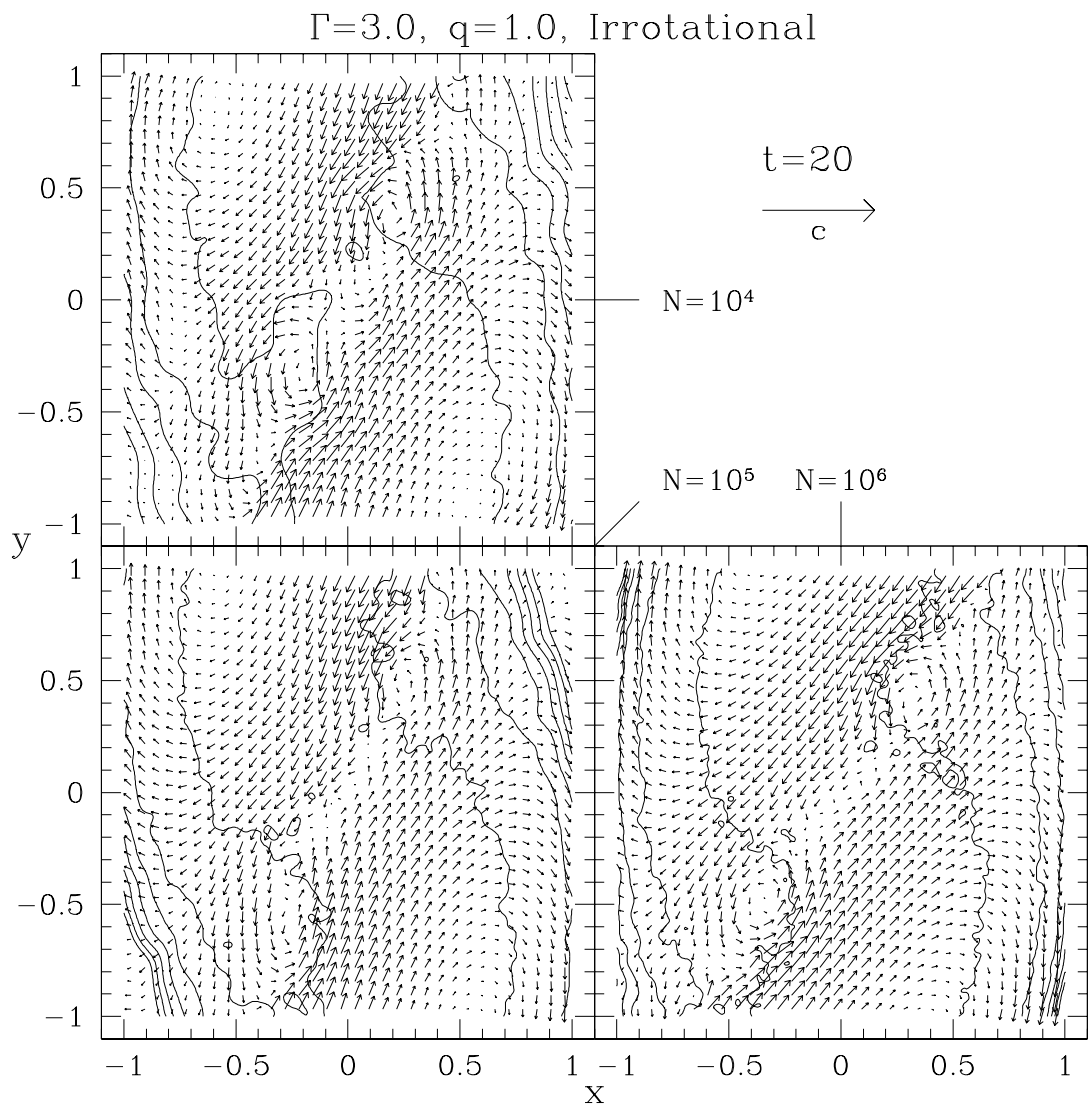}
\caption{Density contours in the orbital plane and velocity field in the
corotating frame of the binary  at
$t=20$ for runs
T1, T2, and T3, shown in Fig.~\protect\ref{fig:gwnp}.
We see some discrepancies in the extent of vortex formation
between the lowest resolution run and the two
higher resolution runs, which agree well.} 
\label{fig:dvnp1}
\end{figure}

\begin{figure}
\centering \leavevmode \epsfxsize=6in \epsfbox{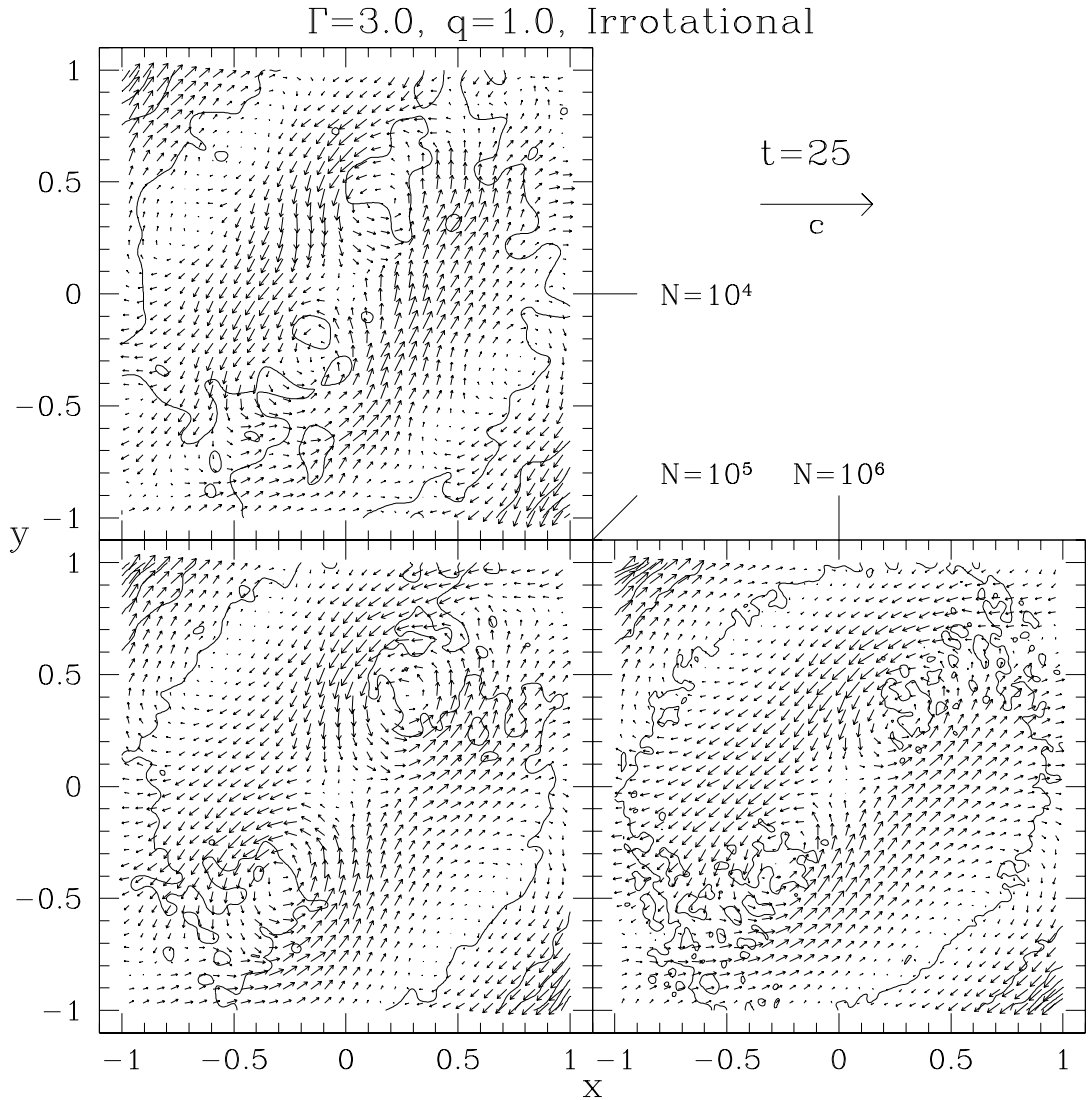}
\caption{Density contours in the orbital plane and velocity field in the
corotating frame of the binary at
$t=25$ for the runs shown in Fig.~\protect\ref{fig:dvnp1}, with the
same conventions as in that figure.  We see that matter between the large
vortices is directed toward the center of the remnant in the two lower
resolution runs, whereas the flow lines are straighter
along the vortex sheet in the highest resolution run.}
\label{fig:dvnp2}
\end{figure}

The vortices forming at the surface of contact are
shown in detail in Figs.~\ref{fig:dvnp1} and \ref{fig:dvnp2}.   
Density contours in the orbital plane are
overlaid with velocity vectors, which are plotted in the corotating
frame of the binary, as defined by Eq.~\ref{eq:omegac}.  
The upper left panels show the evolution of run T1, 
with the bottom left and right panels representing runs T2 and T3, 
respectively. In Fig.~\ref{fig:dvnp1}, we show the state of the three
runs at $t=20$. 
Immediately apparent is that vortices have formed to the largest
extent in the lowest
resolution run, whereas in the higher resolution run there is
little sign of particles mixing, except at a large distance along the
vortex sheet from the
center of the newly forming remnant.  By $t=25$,
shown in Fig.~\ref{fig:dvnp2}, we see that
there is a slight difference between the high
resolution calculations with regard to the direction of the material
flowing along the vortex sheet.  In the highest resolution run, the
streams of material flow nearly in a straight line from one vortex to
the other, whereas in the middle and lowest resolution runs, 
there is a larger region of
material which is accelerated toward the very center of the remnant.
Overall, though, there is excellent agreement between the two highest
resolution calculations.
The agreement of the GW signals calculated from
runs T2 and T3 lead us to believe that the small-scale differences
seen in the matter near the vortex sheet does not carry over
into the bulk of the mass, responsible for the GW
emission.  Essentially, the quadrupole moment at any given instant is
most sensitively dependent upon the orientation of
the densest regions at the cores of the respective NS, which are
unaffected by the small-scale motion in the vortex sheet.  The infall
of the cores is driven by dynamical instability, leading them to
plunge inward and merge, disrupting the vortex sheet and leading to
the formation of the merger remnant.

\section{Gravitational Radiation Spectra}

Zhuge, Centrella, and McMillan first pointed out the importance
of GW energy spectra for the interpretation of merger
signals \cite{Zhu}. In particular, on the basis of Newtonian SPH
calculations, they showed how the observation of particular features 
in the spectra could directly constrain the NS radii and EOS.
Since the detection of NS merger
events will likely be made in narrow-band interferometers, 
it is especially important to
understand the frequency dependence of the GW signals, and not just
their time behavior (as represented by waveforms).  

Following the approach in \cite{Zhu}, we calculate
the GW energy spectrum from each of our calculations as follows.
We first take the Fourier transforms of both polarizations of the GW
signal, 
\begin{eqnarray}
\tilde{h}_+(f)=\int e^{2\pi ift}h_+(t)~dt, \\
\tilde{h}_{\times}(f)=\int e^{2\pi ift}h_{\times}(t)~dt, 
\end{eqnarray}
and we insert them into the following expression giving
the energy loss per unit frequency interval (see, e.g., \cite{Tho}), 
\begin{equation}
\frac{dE}{df}=\frac{c^3}{G}\frac{\pi}{2}(4\pi r^2)f^2\left\langle
|\tilde{h}_{+}(f)|^2+|\tilde{h}_{\times}(f)|^2\right\rangle,
\end{equation}
where the averages are taken over time as well as solid angle.  In
terms of the components of the quadrupole tensor, we then find
\begin{eqnarray}
\frac{dE}{df}&=&\frac{\pi^2 G}{c^5}
\left[\frac{8}{15}\left(|\tilde{Q}^{(2)}_{xx}-\tilde{Q}^{(2)}_{yy}|^2+
|\tilde{Q}^{(2)}_{xx}-\tilde{Q}^{(2)}_{zz}|^2+
|\tilde{Q}^{(2)}_{yy}-\tilde{Q}^{(2)}_{zz}|^2\right)+\right. \nonumber\\
& &\;\;\;\;\;\;\;\;\;
\left.\frac{48}{15}\left(|\tilde{Q}^{(2)}_{xx}|^2+|\tilde{Q}^{(2)}_{yy}|^2+
|\tilde{Q}^{(2)}_{zz}|^2\right)\right],
\label{eq:ps}
\end{eqnarray}
where $\tilde{Q}^{(2)}_{ij}$ represents the Fourier transform of the
second derivative of the traceless quadrupole tensor.

For point-mass inspiral, the energy spectrum takes the power-law form 
$dE/df\propto f^{-1/3}$ \cite{Tho}, the slow {\it decrease\/} with increasing
frequency coming from the acceleration of the orbital decay: although more
energy is emitted per cycle, fewer cycles are spent in any particular
frequency interval as the frequency sweeps up. Near the final merger,
large deviations from this simple power-law spectrum are expected.
However, our initial binary configurations are still reasonably
described by a point-mass model. Therefore, to construct the complete GW signal,
we attach a point-mass inspiral waveform (hereafter referred to
as the inspiral subcomponent) onto the beginning of the signal
calculated numerically with SPH (referred to as the merger
subcomponent). 
The quadrupole tensor for the inspiral subcomponent is assumed to have the
form 
\begin{eqnarray}
Q_{xx}(t')=-Q_{yy}(t')&=&A(t')\cos(\phi(t')) \label{eq:qxx}\\
Q_{xy}(t')&=&A(t')\sin(\phi(t')) \label{eq:qxy}\\
Q_{zz}(t')=Q_{xz}(t')=Q_{yz}(t')&=&0, \label{eq:qxz}
\end{eqnarray}
where $t'\equiv t-t_0$ is the time {\it before\/} our dynamical
calculation starts, and
the amplitude and phase given by
\begin{eqnarray}
A(t')&=&(1+\epsilon)\frac{2M\mu}{[r_0(1-\frac{t'}{t_{\rm mgr}})^{0.25}]}   \\
\phi(t')&=&-\int_{t'}^0 \omega(t') dt'\\
\omega(t')&=&(1.0+\epsilon')\sqrt{\frac{M}{[r_0(1-\frac{t'}{t_{\rm mgr}})^{0.25}]^3}},
\end{eqnarray}
where $r_0$ is the initial binary separation and 
$M$ and $\mu$ are the total and reduced masses of the system,
respectively.  The time constant $t_{\rm mgr}$ is given by the familiar
expression
\begin{equation}
t_{\rm mgr}=\frac{5}{256}\frac{c^5}{G^3}\frac{r_0^4}{\mu M^2}.  \label{eq:t0}
\end{equation}
With $\epsilon=\epsilon'=0$, these expressions correspond to the
well-known quasi-Newtonian point-mass inspiral results\cite{Insp}.
Here the correction factor $\epsilon$ is used to account for
both finite-size and PN effects
and is determined by matching the amplitude of the point-mass signal
to the initial amplitude calculated from our SPH initial condition.
Typically, it is no larger than about $3\%$.  The correction factor
$\epsilon'$ is used in a similar way to match the initial angular
velocity of the system, and is of similar magnitude.  With these
corrections, Eqs.~\ref{eq:qxx}-\ref{eq:qxz} describe well the
inspiral subcomponent at late times, but not, of course, at earlier
times where one should have $\epsilon\rightarrow 0$ and
$\epsilon'\rightarrow 0$ as $t\rightarrow -\infty$.  The discrepancy,
however, is no more than $5\%$, and affects only the nearly
featureless low-frequency part of the spectrum.

\begin{figure}
\centering \leavevmode \epsfxsize=6in \epsfbox{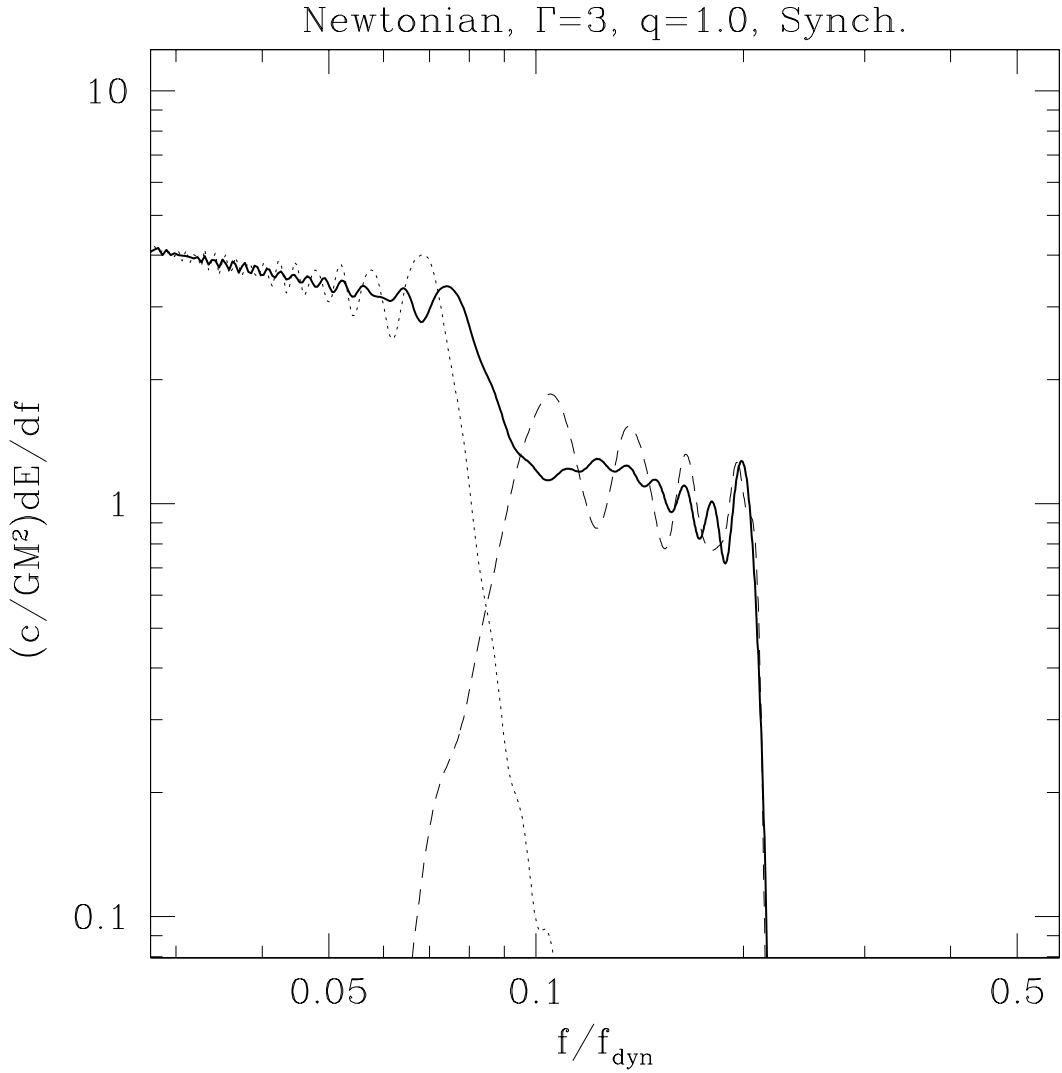}
\caption{Energy spectrum in GW calculated using
Eq.~\protect\ref{eq:ps} from a typical Newtonian calculation
(synchronized, $q=1$, $\Gamma=3$). We show the inspiral (dotted line)
and merger (dashed line) subcomponents of the spectrum, as well as the 
total combined spectrum (heavy solid line).}
\label{fig:psn}
\end{figure}

\begin{figure}
\centering \leavevmode \epsfxsize=6in \epsfbox{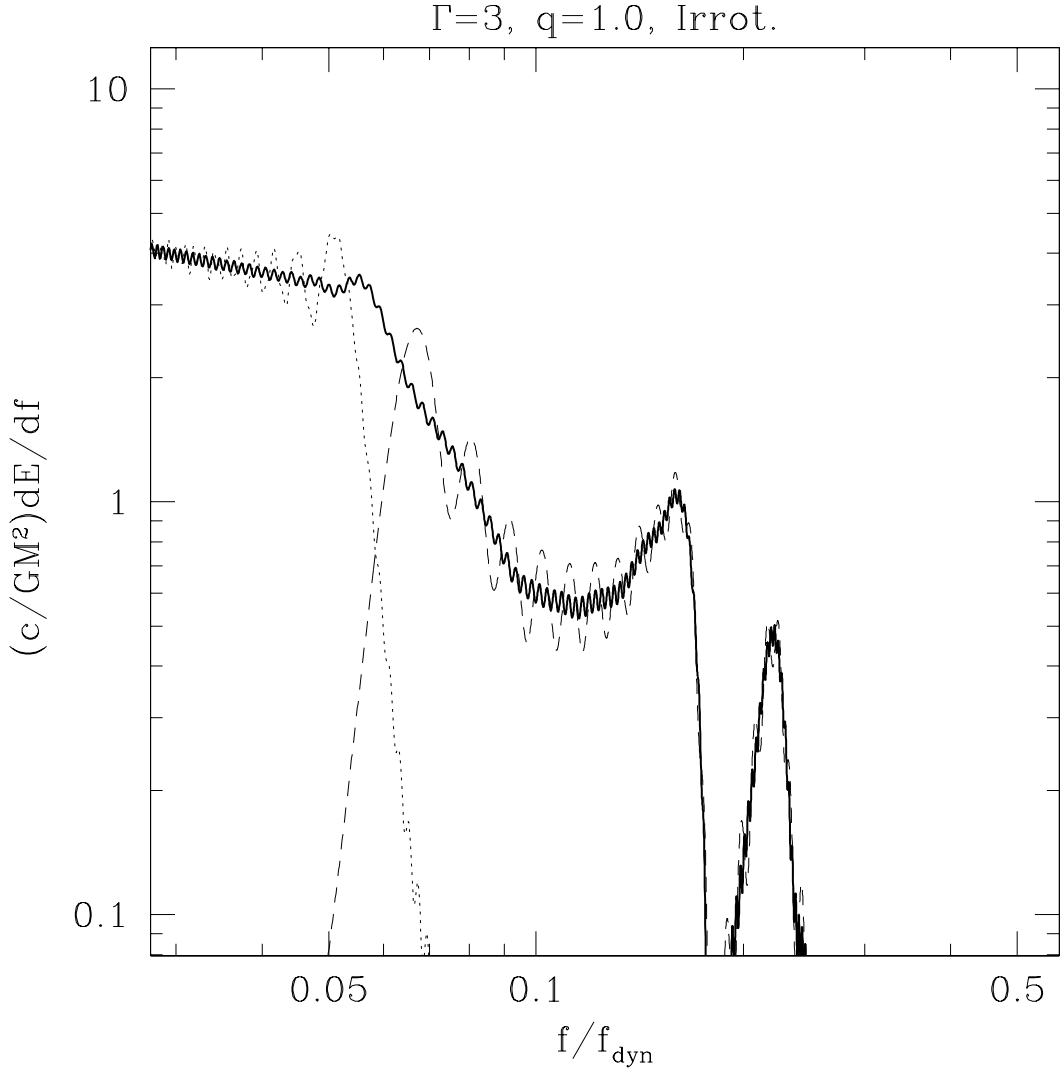}
\caption{Energy spectrum from a typical PN calculation, run E1
($q=1.0$, $\Gamma=3$).
Conventions are as in Fig.~\protect\ref{fig:psn}.  We see that the
energy emitted during the late stages of inspiral is greatly suppressed
relative to the Newtonian case.
The two prominent peaks at higher frequencies correspond to the
maximum GW luminosity and final merger remnant oscillation.} 
\label{fig:psp}
\end{figure}

In Figs.~\ref{fig:psn} and \ref{fig:psp}, we show a comparison between
the energy spectra
computed from a Newtonian calculation with radiation reaction
and from a PN calculation.  The Newtonian calculation 
is the N run from Paper~1 (referred to in Paper~2 as run A1), which
was performed for equal-mass NS with 
$\Gamma=3$ EOS, started at a separation $r_0=3.1$ in a
synchronized initial state.  The PN calculation is the run E1 described in
Sec.~\ref{sec:sep0}, with equal-mass NS, a $\Gamma=3$ EOS, and an
irrotational initial condition with separation $r_0=4.0$.
In both figures, the dashed and dotted lines
represent the merger and
inspiral subcomponents, respectively, with the heavy solid line
representing the total combined spectrum.

We see immediately that there is a significant difference in the energy
emitted between $f/f_{\rm dyn}\simeq 0.06-0.17$ ($550-1600\,{\rm Hz}$ for
our adopted NS parameters; see Eq.~\ref{eq:fdyn}).  
This is directly attributable to
the earlier dynamical instability and faster inspiral rates found 
in PN calculations.  Since the binary
system spends less time in this frequency interval, the energy
emitted is greatly suppressed by the addition of 1PN effects.  The
characteristic ``cliff frequency,'' at which the energy spectrum
plunges below the point-mass power-law, is the best indicator of the start
of dynamical instability. A measurement of this frequency, combined with
theoretical calculations such as those presented here, would lead directly
to the determination of the NS radii, since all frequencies scale with
$(M/R^3)^{0.5}$ and the destabilizing 1PN effects
scale with $M/R$. For the system considered in Fig.~\ref{fig:psp} this 
``cliff frequency'' is about $f=0.06 f_{\rm dyn}$ ($550\,$Hz), almost within
the reach of broad-band interferometers. 
Had the 1PN effects been taken into account at full 
strength, this frequency would likely have been even lower, suggesting that
deviations from point-mass behavior may even become measurable within the
frequency range of current theoretical point-mass inspiral templates.

Two prominent features appear in the PN spectra at much higher frequencies.
In general, the characteristic frequency of GW emission
during mergers sweeps upward monotonically throughout the evolution.
The sharp peaks at $f=0.17 f_{\rm dyn}$ and $f=0.23 f_{\rm dyn}$ ($f\simeq
1600\,{\rm Hz}$ and 
$f\simeq 2200\,{\rm Hz}$) in Fig.~\ref{fig:psp}
are then clearly seen to result from emission at the time of
the GW luminosity peak and during the remnant oscillation (``ring
down'') phase,
respectively. In contrast to some previous Newtonian results \cite{Zhu}, we
find the amplitude of these peaks to be well below the point-mass
power-law (by a factor of almost 3 and 5, respectively).
No calculation we have performed with our PN formalism has ever produced
energy above the point-mass power-law in any frequency range,
confirming that the PN effects not only accelerate the dynamical
instability of the system, but also cause a suppression of the total
GW emission during the entire merger (as we first showed in Paper~1).
Measurements of these peak frequencies would also provide independent 
constraints
on the EOS, but this will certainly require advanced detectors operated 
in narrow-band mode in order to beat the very high laser shot noise above
$\sim1\,$kHz.

\begin{figure}
\centering \leavevmode \epsfxsize=6in \epsfbox{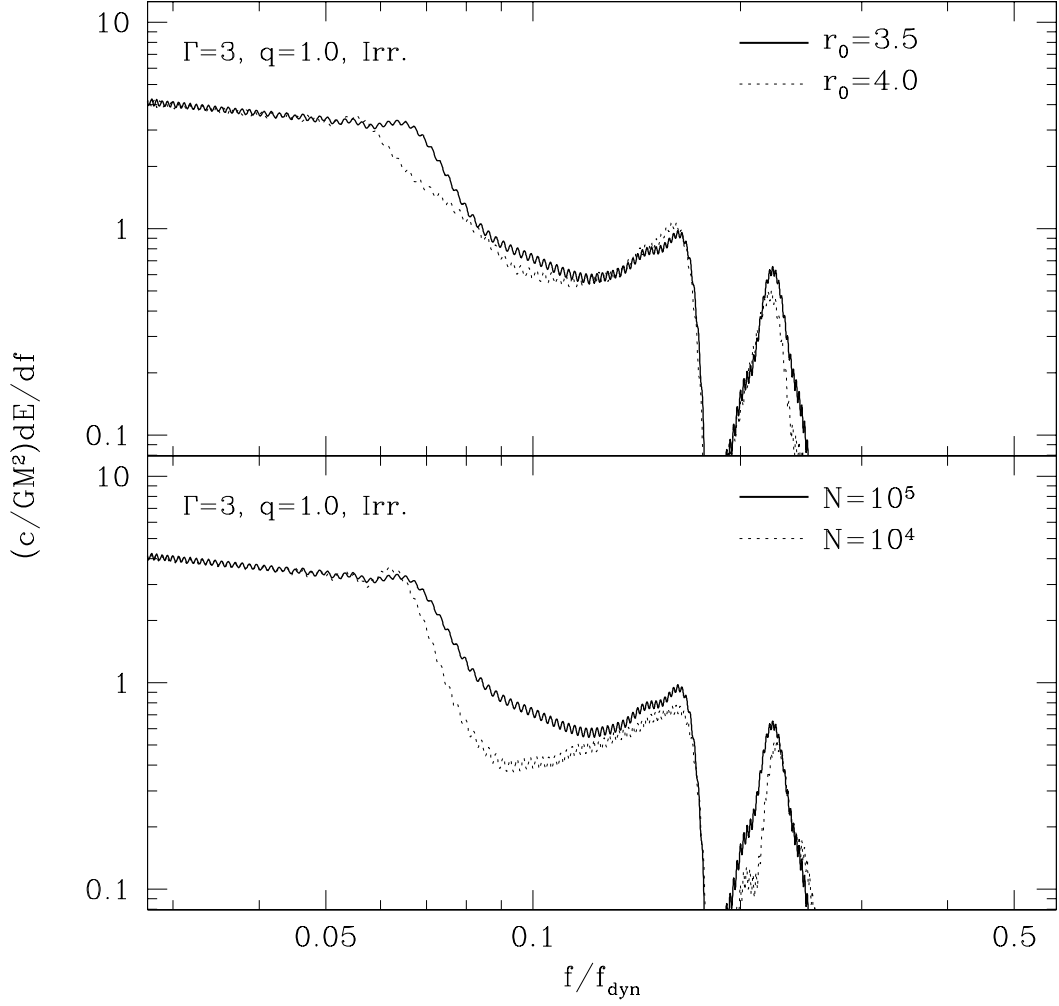}
\caption{Comparison of GW energy spectra calculated from
runs which differ in initial separation (top panel) or
numerical resolution (bottom panel).  In the top panel, we show the
spectra for runs E1 and T2, started from $r_0=4.0$ (thin solid line) and
$r_0/R=3.5$ (thick solid line), respectively. In the bottom panel, we
show the spectra from runs T1 and T2 with $N=10^4$ (thin solid line) and
$N=10^5$ (thick solid line), respectively.}
\label{fig:pssep0}
\end{figure}

We now examine the dependence of these calculated GW energy spectra on initial
binary separation and numerical resolution. In the top panel of 
Fig.~\ref{fig:pssep0}, we compare the spectra computed from our
$\Gamma=3$ EOS, equal-mass calculations started at initial
separations of $r_0=3.5$ (run T2) and $r_0=4.0$ (run E1). 
We see excellent agreement
above $f=0.1 f_{\rm dyn}$ ($\simeq 900\,{\rm Hz}$), but a slight discrepancy
near the cliff frequency. 
At separations in the range $r=3.5-4.0$,
the binary does inspiral faster than a point-mass model would
predict. Since our code has been extensively tested and shown to
reproduce Newtonian results in this separation range (see Fig.~4 of
Paper~1), we attribute the difference to the 1PN effects.

Assuming that a quasi-equilibrium description of the system is appropriate 
in this range of separations, the inspiral rate should be given by
\begin{equation}
\frac{dr}{dt}\simeq \dot{E}_{GW}
\left(\frac{dE}{dr}\right)^{-1}_{equil}
\end{equation}
where $\dot{E}_{GW}$ is the energy loss rate to
gravitational radiation, and $(dE/dr)_{equil}$
is the rate of change of total energy as a function of separation
along an equilibrium sequence of binary NS models \cite{LRS}.  
Since the first factor is very insensitive to PN effects, we
conclude that the slope of the equilibrium energy curve must be made
smaller by the addition of 1PN corrections, in agreement
with the results of PN equilibrium calculations \cite{LomRS}.  While there
must exist a {\it minimum\/} in the equilibrium energy curve at some
critical separation, formally representing the innermost stable circular 
orbit (ISCO) of the system \cite{LRS}, this may be less relevant to 
the onset of dynamical coalescence, which will occur earlier, as soon as
the inspiral timescale becomes comparable to the orbital period.

In the bottom panel of Fig.~\ref{fig:pssep0} we compare the energy
spectra from runs T1 (with $N=10^4$ particles) and T2 ($N=10^5$), 
used to study the dependence of the GW
signal on numerical resolution.  We find that the spectra are
nearly identical in the frequency range $0.15 f_{\rm dyn}<f< 0.27 f_{\rm dyn}$
($1400~{\rm Hz}<f<2500~{\rm Hz}$),
characteristic of the peak emission and the following remnant
oscillations, but disagree by a significant amount at lower
frequencies. Recall from Fig.~\ref{fig:gwnp}
that the lowest resolution run led to larger initial
oscillations around equilibrium.  
While these (spurious) oscillations are at too high a frequency to
appear in the energy spectrum, they do affect the
initial inspiral rate significantly.  The extra energy present in the system, 
combined with the tendency for the binary orbit to become slightly eccentric, 
leads to an artificially accelerated inspiral, and a further reduction of the
energy above the cliff frequency.

\begin{figure}
\centering \leavevmode \epsfxsize=6in \epsfbox{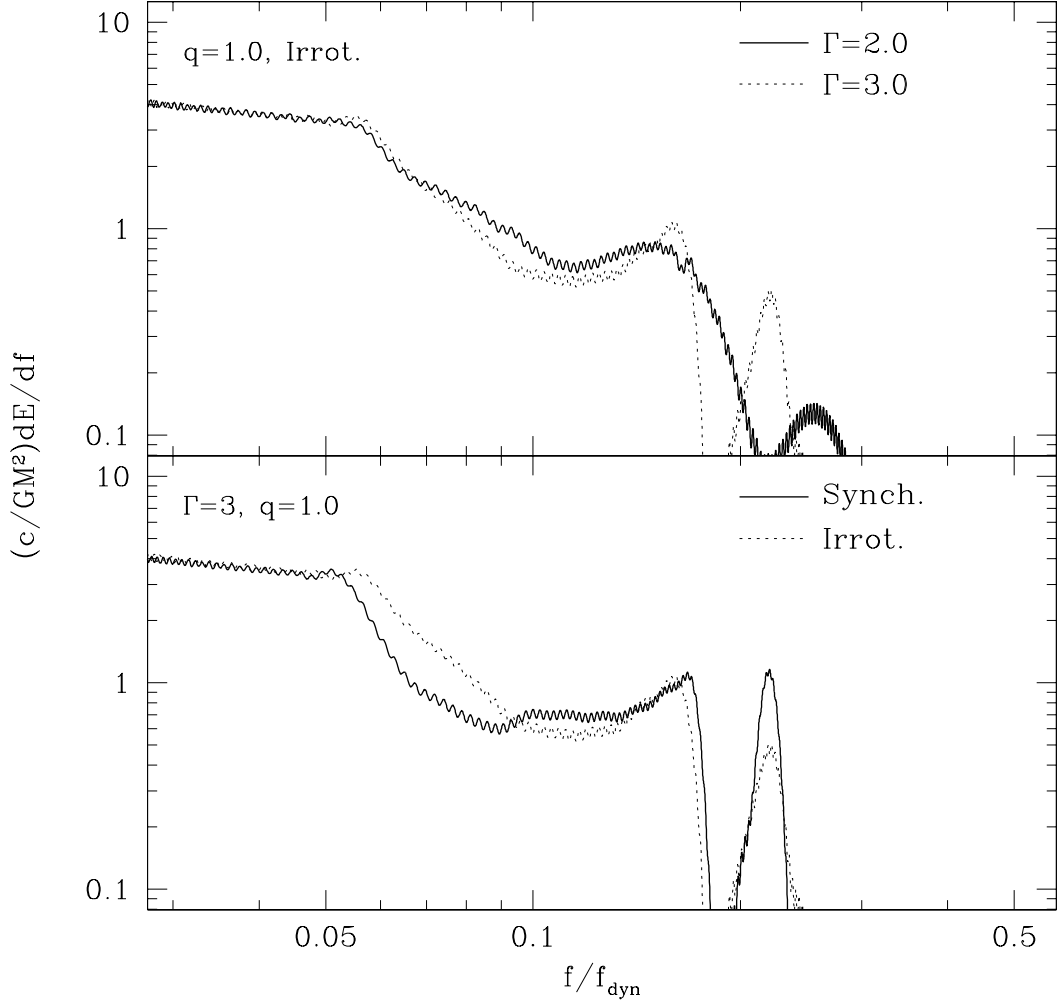}
\caption{Comparison of GW energy spectra from runs which differ in the
choice of EOS (top panel) and initial velocity profile (bottom panel) for equal
mass NS.  In the top panel, we show the spectra for runs E1 and F1,
with a $\Gamma=3$ EOS
(thin solid line) and a $\Gamma=2$ EOS (thick solid line), respectively.
In the bottom panel we show the comparison between run E1 and
run B1 (with a synchronized initial condition; thick solid line).}
\label{fig:pseos}
\end{figure}

Further comparisons are shown in Fig.~\ref{fig:pseos}.
In the top panel, we show the energy spectra for two different values
of $\Gamma$. These were computed from the two
equal-mass runs started from $r_0=4.0$, run E1 with $\Gamma=3$,
and run F1 with $\Gamma=2$. Note that these two different values of $\Gamma$
are compared here at constant $M$ and $R$, which is of course not realistic.
The dominant dependence of the GW emission on the EOS is likely to be 
from the scaling of finite-size effects with $R$, and from the scaling
of PN effects with $M/R$, which we cannot study realistically 
given the limitations of our PN formalism (Sec.~\ref{sec:formal}). At low
frequencies the two spectra are nearly identical.  
For the softer EOS, the peak in the energy spectrum
corresponding to maximum GW emission is of slightly lower amplitude
and frequency.
The peak corresponding to remnant oscillations is greatly suppressed,
since the soft EOS cannot support a stable triaxial configuration (see RS2).
In the bottom panel of Fig.~\ref{fig:pseos}, 
we compare the energy spectra from  synchronized binaries
(run B1) and irrotational binaries (run E1).  We find a sharper
``cliff''  from the synchronized system, leading to lower energy in the
frequency range up to about $f=0.09 f_{\rm dyn}$ ($800\,{\rm Hz}$).  
At higher frequencies, 
however, the two spectra are remarkably similar. In particular, 
the spectral peak frequencies appear to be affected much more strongly 
by the EOS than by the details of the initial velocity configuration.

\begin{figure}
\centering \leavevmode \epsfxsize=6in \epsfbox{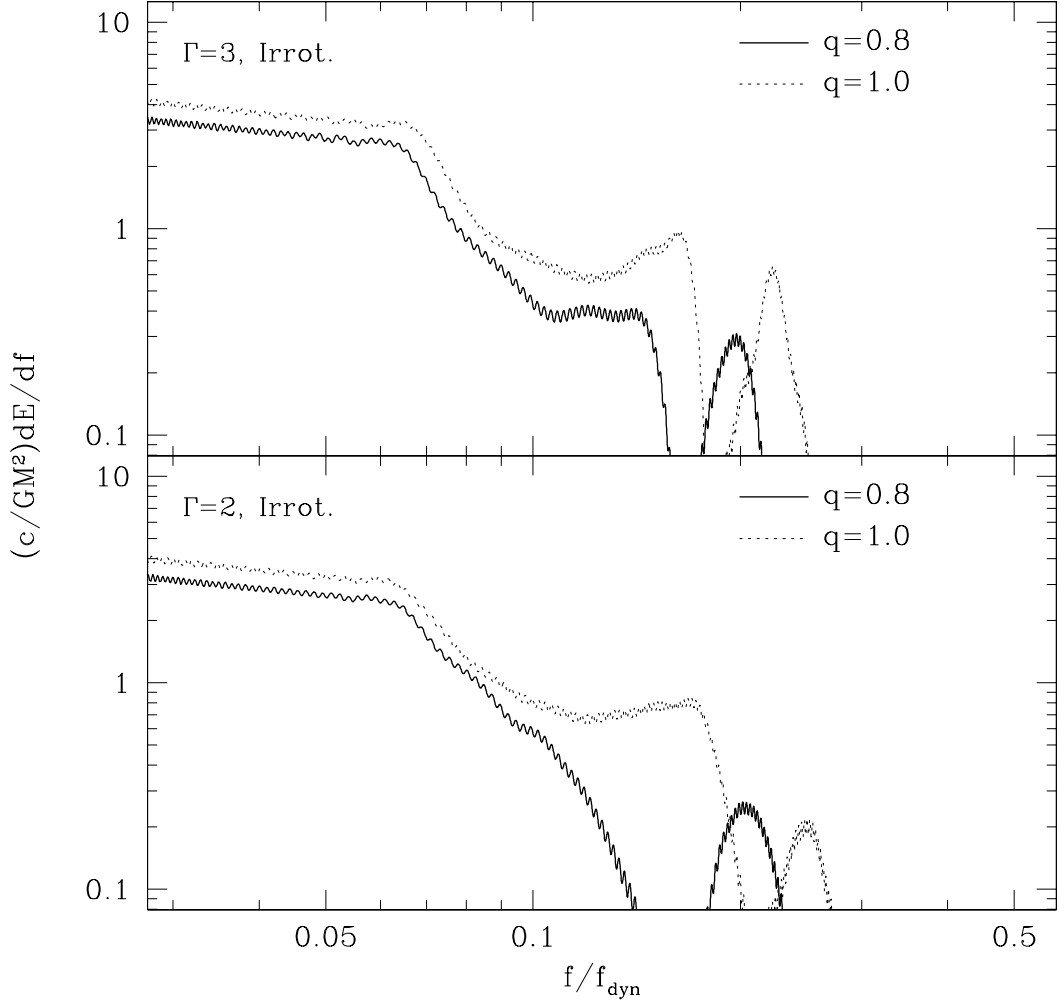}
\caption{Comparison of GW energy spectra for binaries with different
mass ratios and NS with a $\Gamma=3$ (top
panel) or $\Gamma=2$ (bottom panel) EOS.  Thin solid lines represent
the spectra for equal-mass runs T2 and F1, thick solid lines for runs
E2 and F2 with $q=0.8$.  Binaries with $q<1$ not only have
lower amplitude through the lower frequency portion of the spectrum,
but also lack clear peaks at the characteristic frequency of maximum 
GW emission.}
\label{fig:psq8}
\end{figure}

In Fig.~\ref{fig:psq8}, we compare GW spectra for systems with different 
binary mass ratios. In the top panel,
we compare runs T2 and E2, 
with mass ratios of $q=1.0$ and $q=0.8$, respectively (both with a
$\Gamma=3$ EOS and both started from an initial separation 
$r_0=3.5$).  We see a clear difference in the overall amplitude of the
two spectra, but in general the low-frequency behavior of the
spectra is not qualitatively different.  In both cases, we see a
smooth decline in the GW energy which indicates the onset of
dynamical instability, and levels off as we reach the
characteristic frequency of the maximum emission.  A clear
difference is the lack of a true peak in the spectrum for the $q=0.8$
binary, which reflects the suppression of GW emission in
unequal-mass mergers, discussed in Sec.~\ref{sec:q8}.  There is
a well defined peak in the energy spectrum characteristic of emission
from the remnant, nearly equal in amplitude to the first peak, as was
found for the equal-mass binary with the same EOS.  The frequencies
of both peaks in the $q=0.8$ spectrum are shifted lower relative to
their location in the equal-mass case.
In the bottom panel of Fig.~\ref{fig:psq8}, we show the same
comparison for runs F1 and F2 with $\Gamma=2$.
The results are similar except that now the first peak for the
binary with $q=0.8$ is completely absent.

\section{Summary and Directions for Future Work}

Using the PN Lagrangian SPH code described in our previous papers, which is 
complete to 1PN order and includes radiation reaction effects, we have 
investigated a wide parameter space of binary NS mergers started from 
an irrotational initial condition. 
This initial configuration represents the 
most realistic approximation for
NS binaries at separations of $r_0\simeq 3.5-4.0\,R$,
corresponding to the onset of dynamical coalescence.

Based on these calculations, as well as those from our previous papers, we 
have calculated the energy spectrum of the gravitational wave emission 
for a variety of systems.  The key result is the existence of a ``cliff 
frequency'' in all of the energy spectra we have studied, i.e., a 
frequency above which the energy emitted dips dramatically beneath 
the point-mass approximation.  We attribute this effect to the onset of 
dynamical instability in binary systems, which leads to a much more 
rapid inspiral than the point-mass formula predicts, an effect 
amplified when 
PN terms are taken into account.  If the cliff frequencies of binary 
systems are even smaller than those found here (typically around
$500\,{\rm Hz}$ for our standard NS parameters) 
when general relativity is 
treated consistently, they may lie within the frequency band 
accessible to broad-band laser interferometers. For example,
proposals for LIGO II \cite{LIGO2} place its upper frequency limit around 
$1000\, {\rm Hz}$ (where the sensitivity in terms of a characteristic
GW strain has been degraded by a factor $\sim 10$ due to photon shot
noise), meaning that cliff frequencies should be seen if the 
true physical NS radius is sufficiently large, $R\gtrsim 10\, {\rm km}$. 
Our results also suggest that the high frequency features in the GW energy 
spectrum, which result from emission during the merger itself 
($f\simeq 1600\,{\rm Hz}$) and 
from late-time oscillations of the remnant ("ring down" $f\simeq
2200\,{\rm Hz}$),  
will be observable only by more advanced
narrow-band detectors, but if observed, could place strong 
constraints on the NS EOS. 

We also find from our calculations that initially 
irrotational binaries evolve in a qualitatively different way than do 
initially synchronized systems. 
Regardless of the choice of EOS, 
runs started from an irrotational configuration result in much less mass 
shedding than do synchronized runs, depositing no more than $1\%$ of 
the total system mass in an outer halo which remains 
bound to the merger remnant at the center of the system. 
Additionally, there is a significant difference 
in the inspiral rate of such systems immediately prior to merger, with 
synchronized systems merging much more rapidly. 
Thus, since real NS binaries should be essentially irrotational, 
it is important to exercise caution when interpreting the results 
drawn from calculations for initially synchronized systems, 
both before and after the merger occurs. 
The lack of mass shedding seen in our PN calculations 
of irrotational systems 
leads us to conclude that essentially no mass should be shed from 
realistic binary systems, since general relativistic effects further 
suppress the mass shedding instability \cite{Oech2,ShU}. 

From a calculational standpoint, we find competing arguments for the 
ideal initial binary separation of an irrotational system. 
Runs started from larger separations (here $r_0=4.0R$) 
require greater computational resources, and show more spurious 
synchronization because of the numerical viscosity of the SPH method. 
Such calculations do provide a better treatment of the deviations from 
point-mass inspiral prior to merger which result from both Newtonian 
and PN finite-size effects. 
However, runs started from closer in ($r_0=3.5R$) are 
more reliable for drawing conclusions about the final state of the 
system, with regard to mass shedding as well as the rotational 
profile of the remnant.  We find, reassuringly, that the phase and
amplitude evolution of the GW signal during the primary luminosity
peak is unaffected by the initial separation.

We believe that 
our results are unaffected by the limited numerical resolution
of 3D calculations. 
Although merging binary NS systems develop small-scale instabilities, whose 
evolution we cannot follow exactly, we see little effect on the 
GW signals we compute, so long as we use a sufficiently large number 
of SPH particles.  This is especially true for the phase of the GW signal. 
We conclude  
that numerical convergence for a given set of initial conditions and 
physical assumptions is possible without requiring excessive 
computational resources. 

Based on these results, we believe that the fundamental limits of our 
method are not set by the numerical resolution or available 
computational resources, but rather by shortcomings in the PN formalism 
itself.  All our PN calculations are limited by the magnitude of the 
1PN terms that appear in the hydrodynamic equations.  Since we cannot 
treat physically realistic NS models, we deal with the 1PN terms at 
reduced strength.  This hybrid method does in some sense approximate 
the cancellations found in GR between 1PN and higher-order terms, but 
in the end cannot fully model the non-linear nature of relativistic 
gravity.  To do so in a more complete manner, two different approaches 
have been employed.  The first is to attempt to
integrate the hydrodynamics 
equations in full GR, as done recently by Shibata and Uryu \cite{ShU}. 
Starting from quasi-equilibrium irrotational
binary systems at the moment of first contact, they calculate the 
fully relativistic evolution of the system.  Such calculations 
represent a remarkable step forward, and can be considered the 
forefront of the current efforts to understand binary NS coalescence. 
They are limited only by the accuracy to which they can prepare their
initial conditions and by the difficulty
of extracting accurate GW signals from the boundary of 3D grids at
a finite distance, usually well within the near zone of the source.  
As of yet no formalism has proven stable enough to handle a 
calculation which starts from the dynamically stable region and ends 
after the merger is complete. 

A second approach is to use an approximation of full GR
which is known to be numerically stable, known as the {\it 
conformally flat\/} (CF) 
approximation \cite{Oech2,MMW,Bau}. 
In this approximation to general 
relativity, assuming a specific form for the spatial part of the 
metric allows the equations of GR to be reduced to a set of linked 
non-linear {\it elliptic\/} equations.  The drawback to the method is 
that the CF approximation is time-symmetric, and thus does not include 
the dissipative gravitational wave effects seen in full GR.  Such 
terms can be added externally to the formalism, though, to give the 
proper dynamical behavior \cite{Oech2}. 
The CF approximation has been 
used successfully to compute quasi-equilibrium binary sequences 
to high accuracy for both synchronized \cite{Phil,Phil2} and irrotational 
\cite{Phil,Phil2,Irr} binaries, showing good agreement with fully
relativistic calculations of synchronized \cite{RelSyn} and
irrotational binaries \cite{RelIrr}, except for the
case of extremely compact NS at small separations.
Additionally, a PN variant of the CF approximation has been used 
to study rapidly rotating 
single-star configurations \cite{DiffRot}, 
giving excellent agreement with fully GR calculations 
\cite{GRDiffRot}. 
In addition to equilibrium studies, dynamical SPH calculations have
recently been performed using relaxed, 
initially synchronized binary configurations \cite{Oech2}. 

The authors are currently working on a code that will take as an 
initial condition fully relativistic
irrotational, quasi-equilibrium models for binary NS systems, 
calculated using spectral methods \cite{Phil}.  The system will then 
be evolved 
in the CF approximation.  As the formalism is known to work for both 
binary configurations and rapidly rotating single-star configurations, 
reflective of our merger remnants, the hope is that such a method will 
allow us to calculate the evolution of a physically 
realistic binary system from the dynamically stable regime through 
merger and formation of a merger remnant, in a way that is 
consistent with GR throughout.  The comparison of such calculations 
with those performed in full GR will serve as an important check of 
the shortcomings and successes of either approach.

\acknowledgements

The authors wish to thank Thomas Baumgarte, Joan Centrella, Roland
Oechslin, Stu Shapiro and Kip Thorne 
for useful discussions. This work was supported by NSF grants PHY-0070918
and PHY-0133425. Our computational work was
supported by the National Computational Science Alliance under grant 
AST980014N and utilized the NCSA SGI/Cray Origin2000.

\newpage

\newpage
\begin{table}
\caption{Input parameters for the runs described in
this paper: $\Gamma$ is the adiabatic exponent, $q$ the binary mass
ratio, and $r_0$ the initial separation.  All
runs used $10^5$ SPH particles, except testing runs T1 and
T3, with $10^4$ and $10^6$ SPH particles, respectively.  Testing run
T2 differs from run E1 only in the initial separation.
All runs in this paper were started
from an irrotational initial condition.}
\label{table:runs}
\begin{tabular}{c|cccc}
Run & $\Gamma$ & $q$ & $r_0$ & $\log_{10} N$ \\
\tableline\tableline
E1 & 3.0 & 1.00 & 4.0 & 5 \\
E2 & 3.0 & 0.80 & 3.5 & 5 \\ 
\tableline
F1 & 2.0 & 1.00 & 4.0 & 5 \\
F2 & 2.0 & 0.80 & 3.5 & 5 \\ 
\tableline
T1 & 3.0 & 1.00 & 3.5 & 4 \\
T2 & 3.0 & 1.00 & 3.5 & 5 \\
T3 & 3.0 & 1.00 & 3.5 & 6 \\
\end{tabular}
\end{table}

\begin{table}
\caption{Selected results from each run.
Here, $t_0$ is the time at which the run was 
started, and $\theta_{\rm lag}$ is the lag
angle at first contact, given for
the primary and secondary, respectively, for systems with $q\ne 1.0$.
Quantities involving the first
and second GW luminosity peaks are labeled with superscripts
$(1)$ and $(2)$.}
\label{table:gw} 

\begin{tabular}{c|cc|cc|ccc}
Run & $t_0$ & $\theta_{lag} (deg)$ & $L_{\rm max}^{(1)}$ & 
$h_{\rm max}^{(1)}$ & $L_{\rm max}^{(2)}$ & $h_{\rm max}^{(2)}$ & $t^{(2)}$ \\
\tableline\tableline
E1 & -86 & 7.5 & 0.374 & 2.023 & 0.093 & 0.807 & 33 \\
E2 & -39 & 6.5, 8.4 & 0.156 & 1.524 & 0.045 & 0.602 & 35 \\
\tableline
F1 & -105 & 10.0 & 0.479 & 2.129 & 0.050 & 0.545 & 30 \\
F2 & -55 &  3.8, 12.1 & 0.098 & 1.364 & 0.056 & 0.564 & 35 \\
\tableline
T1 & -21 & 7.8 & 0.358 & 2.009 & 0.125 & 0.932 & 32 \\
T2 & -3 & 12.0 & 0.337 & 1.972 & 0.087 & 0.746 & 30 \\
T3 & -16 & 5.8 & 0.356 & 1.989 & 0.111 & 0.907 & 32 \\
\end{tabular}
\end{table}

\begin{table}
\caption{Properties of the merger remnants.
Here $M_r$ is the rest mass of the
remnant, $M_{gr}$ is its gravitational mass, $a_r$ is its Kerr parameter,
$\Omega_c$ and $\Omega_{eq}$ are the angular rotation
velocities at the center and at the equator, and the $a_i$'s and
$I_i$'s are the radii of the principal axes and moments of inertia.}
\label{table:final} 

\begin{tabular}{c|ccccccccc}
run & $M_r$ & $M_{gr}$ & $a_r$ & $\Omega_c$ & $\Omega_{eq}$ & $a_1$ &
$a_2/a_1$ & $a_3/a_1$ & $I_2/I_1$ \\
\tableline\tableline
E1 & 1.95 & 1.88 & 0.73 & 0.668 & 0.435 & 1.81 & 0.94 & 0.55 & 1.07 \\
F1 & 1.94 & 1.79 & 0.81 & 0.838 & 0.436 & 1.83 & 0.96 & 0.49 & 1.01 \\
T1 & 1.97 & 1.91 & 0.74 & 0.711 & 0.433 & 1.81 & 0.94 & 0.54 & 1.10 \\
T2 & 1.97 & 1.91 & 0.74 & 0.735 & 0.414 & 1.80 & 0.98 & 0.57 & 1.07 \\
\end{tabular}
\end{table}

\begin{table}
\caption{GW quantities computed for runs with
different numbers of particles $N$ at representative times.  Here,
$h(t)$ is the GW strain, 
and $\Omega_{GW}(t)$ is the instantaneous angular frequency of the
GWs.  At $t=10$, the stars are about to make contact, $t=20$
is the moment of peak GW luminosity, and by $t=30$
a remnant has begun to form.  In all cases, we see much 
better agreement between
the two higher resolution runs, at a level of $\simeq 2\%$. } 
\label{table:gravnp}

\begin{tabular}{c|ccccccc}
run & N & $h(t=10)$ & $\Omega_{GW}(t=10)$ & $h(t=20)$ & $\Omega_{GW}(t=20)$ &
$h(t=30)$ & $\Omega_{GW}(t=30)$ \\  
\tableline\tableline
T1 & $10^4$ & 1.49 & 0.556 & 1.71 & 1.053 & 0.74 & 1.192 \\
T2 & $10^5$ & 1.57 & 0.584 & 1.91 & 0.978 & 0.90 & 1.166 \\
T3 & $10^6$ & 1.58 & 0.581 & 1.87 & 0.996 & 0.89 & 1.154 \\
\end{tabular}
\end{table}

\end{document}